\documentclass[10pt]{article}
\usepackage{amsmath,amsfonts,amssymb,amsthm,bbm,bm}
\usepackage{pdfsync}
\usepackage{float}
\usepackage{subfigure}
\usepackage[titletoc]{appendix}
\usepackage{graphicx}
\usepackage[latin1]{inputenc}
\usepackage{hyperref}
\usepackage[all]{xy}
\usepackage{stmaryrd}
\usepackage{psfrag}
\usepackage{rotating}
\usepackage{epstopdf}
\usepackage{braket}
\usepackage{amsthm}

\newcommand{\bit}{\begin{itemize}}
\newcommand{\eit}{\end{itemize}}
\newcommand{\bd}{\begin{description}}
\newcommand{\ed}{\end{description}}

\newcommand{\bc}{\begin{center}}
\newcommand{\ec}{\end{center}}

\newcommand{\be}{\begin{equation}}
\newcommand{\ee}{\end{equation}}
\newcommand{\bea}{\begin{eqnarray}}
\newcommand{\eea}{\end{eqnarray}}
\newcommand{\bs}{\begin{subequations}}
\newcommand{\es}{\end{subequations}}

\begin{document}

\title{\bf Bianchi I effective dynamics in Quantum Reduced Loop Gravity}
\author{\Large{Emanuele Alesci$^a$, Gioele Botta$^b$, Giovanni Luzi$^c$ and Gabriele V. Stagno$^{d,e}$}
	\smallskip \\ \small{$^{a}$Institute for Gravitation and the Cosmos, Penn State, University Park, PA 16802, U.S.A} \\  
	\small{$^b$Faculty of physics, University of Warsaw, Pasteura 5, 02-093 Warsaw, Poland}\\
	\small{$^c$Institute of Fluid Mechanics, FAU Busan Campus, University of Erlangen-Nuremberg, 46742 Busan, Republic of Korea}\\
	\small{$^d$Sapienza University of Rome, P.le Aldo Moro 5, (00185) Roma, Italy }\\
	\small{$^e$Aix Marseille Univ., Univ. de Toulon, CNRS, CPT, UMR 7332, 13288 Marseille, France}
}
\date{\today}

\maketitle
%%%%%%%%%%%%%%%%%%%%%%%%%%%%%%%%%%%%%%%%%%%%%%%%%%%%%%%%%%%%%%%%%%%%%%%%%%%%%%%%%%
\begin{abstract}
	\noindent
The	effective quantum dynamics of Bianchi I spacetime is addressed within the statistical regularization scheme in Quantum Reduced Loop Gravity. The case of a minimally coupled massless scalar field is studied and compared with the effective $\bar{\mu}-$Loop Quantum Cosmology. The dynamics provided by the two approaches match in the semiclassical limit but differ significantly after the bounces. Analytical and numerical inspections show that energy density, expansion scalar and shear are bounded also in Quantum Reduced Loop Gravity and the classical singularity is resolved for generic initial conditions in all spatial directions.
\end{abstract}

\maketitle
\tableofcontents 

%%%%%%%%%%%%%%%%%%%%%%%%%%%%%%%%%%%%%%%%%%%%%%%%%%%%%%%%%%%%%%%%%%%%%%%%%%%%%%%%%%%
\section{Introduction}

Quantum Reduced Loop Gravity (QRLG) \cite{Alesci:2013xd,Alesci:2016gub} is aimed to address symmetric sectors of Loop Quantum Gravity (LQG) \cite{Rovelli:2004tv,Thiemann:2007zz} and it has proved to be a versatile and powerful tool for both primordial cosmology \cite{Alesci:2016xqa,Alesci:2017kzc} and black hole physics \cite{Alesci:2018loi}. It is based on suitable gauge fixings of LQG (see e.g. \cite{Alesci:2018ewg}) and reduces the computational task that plague the full theory whilst retaining its main features - graph and intertwiner structure - allowing a deeper theoretical understanding \cite{Alesci:2014uha,Alesci:2016rmn} and the actual computation of observational consequences \cite{Alesci:2018qtm}.

QRLG has been originally designed for dealing with cosmology, and here it has been successful in bridging Loop Quantum Cosmology (LQC) (see \cite{Agullo:2016tjh} for a recent review) to full LQG \cite{Alesci:2016rmn,Alesci:2014rra}. From the QRLG perspective, LQC stands as a first order effective quantization that can be refined within QRLG including key quantum terms coming from the full theory. Those corrections are crucial and they must be taken into account when one is interested in questioning the deep quantum regime of the universe. For instance, for the Friedmann Lemaitre Robertson Walker (FLRW) model they provide a quantum evolution which significantly differs from the one given by LQC: the Big Bounce scenario provided by the former \cite{Ashtekar:2006wn} is replaced in QRLG by the dynamics of an emergent-bouncing universe \cite{Alesci:2016xqa}. The universe ``emerges" from an infinite past with a non-vanishing volume that it keeps until a transient phase of expansions and contractions occurs and eventually matches the LQC evolution. This alternative cosmology has been recently considered also in different contexts (see \cite{Brandenberger:2017pjz} and references there in) and its observational signatures have been studied in \cite{Martineau:2018isp}, and by some of the authors in \cite{Alesci:2018qtm,Olmedo:2018ohq}. 

Going beyond the isotropic context is a needed step both for testing the QRLG approach in a more general setting and for addressing the issue of isotropization (e.g. see \cite{Cianfrani:2010gj}), a mechanism that is believed to have a quantum origin and to be responsible for the observed large scale symmetry of our universe. In this paper we study the homogeneous anisotropic sector of QRLG associated to the Bianchi I geometry, within the statistical regularization scheme \cite{Alesci:2017kzc}. This framework provides an effective graph-changing dynamics for both isotropic and not isotropic sectors and includes LQC regularizations as special cases. An effective Hamiltonian for the (quantum corrected) geometry of the Bianchi I universe is considered by taking the expectation value of the (reduced) non-graph-changing Hamiltonian operator over a gaussian ensemble of coherent states peaked on classical Bianchi I phase-space and based on cuboidal graphs with different number of nodes $A_i$. For large $A_i\,$, one can expand the integral over the ensemble that defines the QRLG-effective Hamiltonian, finding a zero order contribution which concides with the standard LQC Bianchi I effective Hamiltonian plus (infinitely many) corrections that become relevant in the deep quantum era \cite{Alesci:2017kzc}. Here we have addressed the dynamics of the QRLG model considering all those contributions, i.e. without approximating the integral to a given order, and  made a comparison with the dynamics provided by the standard effective LQC in the presence of a massless sclar field $\phi$. The usual Hamilton eqs. are used to obtain the associated effective dynamics and the evolution is numerically studied for some general initial conditions, i.e. isotropic, ``Kasner-like" (one direction expands/contracts and the other two contract/expand) and ``un-Kasner like" (all directions expand/contract). Similarly to what happens in the isotropic case\footnote{for both the ``volume counting'' \cite{Alesci:2016xqa} and  ``area counting'' \cite{Alesci:2017kzc} statistical regularizations.}, the QRLG-quantum corrections to the classical Bianchi I model provide a non singular dynamics that significantly differs from the one given by LQC before the bounces and matches it afterwards. More specifically, the QRLG model turns out not to bridge two classical Bianchi I universes, as instead it occurs in LQC \cite{Chiou:2007mg}. Starting from a classical Bianchi I and going backward in the relational time $\phi$, the universe undergoes three bounces (one in each direction) and after that its scale factors start growing faster than they do in LQC and GR. Moreover, our numerical simulations show that the singularity is avoided in all directions and for generic initial conditions. In particular, for Kasner-like initial conditions each directional scale factor is non vanishing, contrary to what happens in the LQC-evolution where one of the scale factor goes to zero in the far past \cite{Chiou:2007mg}. 

The paper is organized as follows. We start recalling useful definitions for the Bianchi I geometry and its Hamiltonian formulation in terms of the Ashtekar variables. The relevant kinematical quantities, such as the directional Hubble rates, the expansion scalar and the shear, are then introduced. In sec.\ref{Sec_Classical_and_LQC} the Hamiltonian constraints for the Bianchi I model are given for both GR and LQC. The QRLG-Bianchi I model is defined in sec.\ref{Sec_QRLG_BianchiI} and the expressions of the kinematical quantities, defined in general terms in sec.\ref{Sec_BianchiIgeo}, are here given explicitely for our QRLG-model. Analytical bounds for those quantities and for the energy density are computed and their evolutions numerically studied in sec.\ref{Sec_Upperbounds}. Section \ref{Sec_effectivedynamics_numerical} presents the numerical study of the QRLG-Bianchi I effective dynamics for the minimally coupled massless scalar field, compared to the one provided by LQC. Finally, the last section is devoted to conclusions and outlooks.

 Throughout the paper we use $\gamma=0.24$ and $G=\hbar=c=1\,,$ so that $l^2_P:=\hbar G/c^3=1$ . We don't follow Einstein notation, i.e. repeated indices are not summed otherwise explicitly said.

%%%%%%%%%%%%%%%%%%%%%%%%%%%%%%%%%%%%%%%%%%%%%%%%%%%%%%%%%%%%%%%%%%%%%%%%%%%%%%%%%%%

\section{Bianchi I geometry and related key quantities}\label{Sec_BianchiIgeo}

We review here some useful definitions for the Bianchi I model in the Ashtekar variables. Everything in this section holds for GR, effective LQC and effective QRLG.

\subsubsection*{Line element}

We choose cartesian comoving coordinates $(t,x^b)$ and unitary lapse function. The Bianchi I geometry is then associated to the following line element:
\begin{equation}
ds^2=-dt^2+\sum_{bc}q_{bc}(t)dx^bdx^c=-dt^2+\sum_b (a_b(t)dx^b)^2\label{dsBianchi}\,,
\end{equation}
where $a_b(t)$ are the scale factors for the spatial directions $b=1,2,3\,.$

\subsubsection*{Simplectic structure}

The Hamiltonian formulation for the geometry (\ref{dsBianchi}) closely follows the one implemented for the FLRW case \cite{Ashtekar:2003hd}. Once a fiducial \textit{cuboidal} cell $\mathcal{V}$ of coordinate sides $\tilde{L}^i$ is introduced, a parametrization of the phase-space associated to the geometry is provided by the Ashtekar variables, i.e. by the connection $A^i_a (t)$ and the densitized triad $E^a_i(t)$ \cite{Ashtekar:2009vc}: 
\be
A^i_a (t):=\frac{c^i(t)}{\tilde{L}^i}\, \tilde{e}^i_a\,,\qquad  E^a_i(t):=\frac{p_i(t)}{\tilde{V}}\tilde{L}^i\, \tilde{e}^a_i\,,\label{Ashtekarvariables}
\ee
where $\tilde{e}^i_a=\delta^i_a$ and $ \tilde{e}^a_i=\delta^a_i$ are, respectively, the orthonormal flat fiducial co-triad and triad field adapted to $\mathcal{V}$, $$q_{bc}=\sum_{ik}a_b\tilde{e}^i_b\,a_c\tilde{e}^k_c\,\delta_{ik}\,,$$
  and $\tilde{L}^j$ are the set of coordinate lenghts defining the coordinate volume $\tilde{V}=\tilde{L}^1 \tilde{L}^2 \tilde{L}^3$ of the fiducial cell, which is related to the physical volume\footnote{In QRLG this quantity is taken to be the volume of the biggest observable region of the universe, as explained in \cite{Alesci:2017kzc}.} $V$ as
 \be
 V=a_1a_2a_3\tilde{V}=\sqrt{p_1 p_2  p_3}\,.
 \ee
 Connections and triads are diagonal matrices whose entries satisfy the following Poisson bracket:
\be
\{c^i,p_j\}=8\pi \gamma \delta^i_j\,, \label{BianchiPoissonBrackets}
\ee
which defines the simplectic structure for the geometrical sector of the model. The $p_j$ are related to the scale factors $a_k$ as follows (we choose a positive orientation) 
\be
p_i=\tilde{L}^j \tilde{L}^k |a_j a_k|\qquad i=1,2,3\quad i\neq j\neq k\,. \label{Bianchitriads}
\ee
Finally, when the geometry is sourced by a scalar field, the phase-space gets enlarged and coordinatized by the 8-tuples $(c^i,p_i,\phi,p_{\phi})$, where 
\be
\{\phi,p_{\phi}\}=1\,.\label{phipoissonbrackets}
\ee

\subsubsection*{Dynamics and energy density}

Hereafter we will refer to the case of the Bianchi I geometry filled with a massless scalar field $\phi$. 

The dynamics is generated by the following Hamiltonian constraint: 
\be
\mathcal{C}(c^i,p_i,p_{\phi}):=H_{BI}(c^i,p_i)+H_{\phi}(p_{\phi},p_i)\approx 0\,,\label{Bianchiconstraint}
\ee
where $H_{BI}:=H_{gr},H_{lqc}, H$ generically refers to the geometrical sector in GR, LQC and QRLG, respectively (for their actual definitions see the following sections), and 
\begin{equation}
H_{\phi}:=\frac{p_{\phi}^2}{2V}\,,\label{Hscalarfield}
\end{equation}
is the kinetic energy of the field $\phi$, the only contribution coming from the matter sector. The Hamilton eqs. follow:
\be
\dot{c}^i=8\pi\gamma\frac{\partial \mathcal{C}}{\partial p_i}\,,\quad \dot{p_i}=-8\pi\gamma\frac{\partial \mathcal{C}}{\partial c^i}\,,\quad\dot{\phi}=\frac{\partial H_{\phi}}{\partial p_{\phi}}\,,\quad \dot{p}_{\phi}=-\frac{\partial H_{\phi}}{\partial \phi}\,.\label{BianchiHeqq}
\ee

Soon we will be interested in evaluating the field energy density $\rho$ along the physical motions. It is given by the ratio
\be
\rho:=\frac{H_{\phi}}{V}\approx-\frac{H_{BI}}{\sqrt{p_1p_2p_3}}\label{rho}\,.
\ee

\subsubsection*{Directional Hubble rates, expansion scalar and shear}

The main kinematical quantities are the directional Hubble rates $H_i:=\dot{a}_i/a_i\,$. In terms of the triads (\ref{Bianchitriads}) they read
\be
H_i=\frac{1}{2}\left(-\frac{\dot{p}_i}{p_i}+\frac{\dot{p}_j}{p_j}+\frac{\dot{p}_k}{p_k}\right)\qquad i=1,2,3,\quad i\neq j \neq k\,,\label{Hubblerates}
\ee
and from them two more useful quantities are built: the expansion scalar $\theta\,$,
\be
\theta:=\frac{1}{V}\frac{dV}{dt}=\sum_i H_i\,,\label{Expansion}
\ee
and the shear $\sigma^2\,,$
\be
\sigma^2:=\sum_i H^2_i-\frac{\theta^2}{3}=\frac{1}{3}[(H_1-H_2)^2+(H_2-H_3)^2+(H_3-H_1)^2]\,,\label{Shear}
\ee
which clearly vanishes in the isotropic limit.

%%%%%%%%%%%%%%%%%%%%%%%%%%%%%%%%%%%%%%%%%%%%%%%%%%%%%%%%%%%%%%%%%%%%%%%%%%%%%%%%%%%%%%%%%%%%%%%%

\section{Classical and effective-LQC constraints for Bianchi I}\label{Sec_Classical_and_LQC}

In the following sections we will compare the dynamics of the QRLG model to the one provided by LQC. For this pourpose and in order to understand the choice of the initial conditions for the dynamical problem, we recall here the Hamiltonian constraints for the Bianchi I geometry in GR and LQC.

The classical Bianchi I universe is associated to the constraint\footnote{hereafter we will deliberately loose track of covariance/contravariance using only downstairs indices.} $\mathcal{C}_{gr}$\,,
\begin{equation}
\mathcal{C}_{gr}:=H_{gr}+H_{\phi}:= -\frac{1}{8\pi \gamma^2}\frac{\left(c_2p_2c_3p_3+c_1p_1c_3p_3+c_1p_1c_2p_2 \right)}{\sqrt{p_1 p_2 p_3}}+\frac{p_{\phi}^2}{2\sqrt{p_1p_2p_3}}\approx 0\,, \label{constraintGR}
\end{equation}
where the $p_i$ follow the general definition (\ref{Bianchitriads}) and the connections are proportional to the directional Hubble rates:
\begin{equation}
c_i=\gamma L_i H_i\label{connectionGR}\,,
\end{equation}
being $L_i:=a_i\tilde{L}_i$. Note that (\ref{connectionGR}) strictly holds in GR and it is only approximately true in LQC and QRLG in the classical limit $p_i>>1,c_i<<1$, where we will choose the initial conditions for the dynamical problem for both LQC and QRLG (see below).

The effective $\bar{\mu}$-LQC of Bianchi I is obtained \cite{Chiou:2007mg, Ashtekar:2009vc} by replacing the classical connections in  \eqref{constraintGR} according to the ``polymeric prescription", i.e.:
\begin{equation}
c_i\rightarrow\frac{\sin(\bar{\mu}_i c_i)}{\bar{\mu}_i}
\end{equation}
where 
\begin{equation}
\bar{\mu}_i:=\sqrt{\Delta_{LQG}}\sqrt{\frac{p_i}{p_j p_k}} \qquad i\neq j\neq k
\end{equation}
and $\Delta_{LQG}=5.22$ is the LQG area gap. The resulting effective constraint $\mathcal{C}_{lqc}$ reads\footnote{We neglect holonomy corrections. Those are expected to be subleading for super-Planckian volumes, condition that is always met in our numerical simulations, see sec.\ref{Sec_effectivedynamics_numerical}.}:
\begin{equation}
\mathcal{C}_{lqc}:=H_{lqc}+H_{\phi}:= -\frac{1}{8\pi \gamma^2\sqrt{p_1 p_2 p_3}}\left(\frac{\sin(\bar{\mu}_2 c_2)\sin(\bar{\mu}_3 c_3)}{\bar{\mu}_2\bar{\mu}_3}p_2p_3 +\mbox{\,cyclic terms}\right)+\frac{p_{\phi}^2}{2\sqrt{p_1p_2p_3}}\approx 0 \,.\label{constraintLQC}
\end{equation}

%%%%%%%%%%%%%%%%%%%%%%%%%%%%%%%%%%%%%%%%%%%%%%%%%%%%%%%%%%%%%%%%%%%%%%%%%%%%%%%%%%%%%%%%%%%%%%%%%%%%%%%%%%%%%%%%%%%%%%%%%%%%%

\section{The QRLG-Bianchi I model }\label{Sec_QRLG_BianchiI}

The effective Hamiltonian we introduce here is the one provided by QRLG within the ``area counting" statistical regularization scheme \cite{Alesci:2017kzc}. Its expression is given by the expectation value of the (reduced) scalar constraint $\hat{H}^R$ over a classical mixture of coherent states based on cuboidal graphs with different number of nodes $A_i$. The mixture is described by the following density matrix:

\begin{equation}
\hat{\rho}_{A}:=\prod_i \sum_{A_i=1}^{A_i^{max}}\left(\begin{matrix} A_i^{max}\\ A_i \end{matrix}\right)|A_i,j_i,\theta_i\rangle \langle A_i,j_i,\theta_i|\,, \label{densitymatrix}
\end{equation}
where 
$|A_i,j_i,\theta_i\rangle$ are the Thiemann's coherent states in the kinematical space of QRLG \cite{Alesci:2014uha}, peaked on both the intrinsic and extrinsic geometry of classical Bianchi I, i.e. on the QRLG fluxes $E_i=8\pi\gamma l_P^2 j_i$ and holonomies $h_l=e^{i\theta_l j_l}$. The maximum number of nodes contained in the physical area $p_i$ is
\be
A_i^{max}=\frac{2p_i}{\Delta'}\label{areamax}
\ee
 where $\Delta'=6.03$ is the ``area counting" area gap in QRLG\footnote{which is slightly greater then the usual LQG-value $\Delta_{LQG}=5.22$, as explained in \cite{Alesci:2017kzc}.} and the expectation value 
\be
H^{disc}:=\frac{Tr(\hat{\rho}_{A}\,\hat{H}^R)}{Tr \hat{\rho}_A}\label{H_qrlg_definition}\ee
explicitely reads:
\be
H^{disc}(\{A^{max}_i(p_i)\},\{c_i\})=-\frac{1}{8\pi \gamma^2} \frac{\left[\prod_i\sum_{1}^{A^{max}_i}\left(\begin{matrix}A^{max}_i \\ A_i \end{matrix}\right) \right]\,\tilde{H}(\{A^{max}_i(p_i)\},\{c_i\};\{A_i\})}{\prod_i\sum_{1}^{A^{max}_i}\left(\begin{matrix}A^{max}_i \\ A_i \end{matrix}\right)}\,,\label{HBianchi_QRLG_disc}
\ee
where
\be
\tilde{H}:= A_1\sqrt{\frac{p_2p_3}{p_1}}\sin\left(c_2\sqrt{\frac{A_2}{A_1A_3}}\right)\sin\left(c_3\sqrt{\frac{A_3}{A_1A_2}}\right)+ \mbox{\,cyclic terms}\,.
\ee
For $A_i>>1$ we will use the continuous approximation\footnote{Note that already for $A_i>12$ we have good agreement with the exact expression \eqref{HBianchi_QRLG_disc} (see figs. \ref{Fig_I1} and \ref{Fig_I2}), thus, in sec.\ref{Sec_effectivedynamics_numerical} the dynamics has been studied within the continuous approximation.} for the binomials considering the following expression 

\be
H(\{p_i\},\{c_i\}):=-\frac{1}{8\pi \gamma^2} \frac{\left[\prod_i\int_{1}^{2p_i/\Delta'}  e^{-\frac{\Delta'}{p_i}(A_i-\frac{p_i}{\Delta'})^2}\,dA_i\right]\,\tilde{H}(\{p_i\},\{c_i\};\{A_i\})}{\prod_i\int_{1}^{2p_i/\Delta'}  e^{-\frac{\Delta'}{p_i}(A_i-\frac{p_i}{\Delta'})^2}\,dA_i}\,.\label{HBianchi_QRLG_contapprox}
\ee
as our Hamiltonian for the geometrical sector. Including the contribution of the matter sector, we find the total constraint
\be
\mathcal{C}_{qrlg}:=H+H_{\phi}\approx 0\,,\label{constraint_Bianchi_QRLG}
\ee
which completes the definition of our model.

%%%%%%%%%%%%%%%%%%%%%%%%%%%%%%%%%%%%%%%%%%%%%%%%%%%%%%%%%%%%%%%%%%%%%%%%%%%%%%%%%%%%%%%%%%%%%%%%%%%%%%%%%%%%%%%%%%%%%%%%%%%
\subsection{$\rho,\theta$ and $\sigma^2$ explicit expressions for the model}

In this section we provide the explicit expressions for the phase-space functions \eqref{rho}, \eqref{Expansion},\eqref{Shear} for the constraint \eqref{constraint_Bianchi_QRLG}. The analytical bounds and the numerical evolutions along physical motions for those quantities, are discussed in sec.\ref{Sec_Upperbounds}.

\subsubsection*{Energy density}
After a sign change, the ratio between \eqref{HBianchi_QRLG_contapprox} and the physical volume V gives the energy density
\be
\rho(\{p_i\},\{c_i\})=\frac{1}{8\pi \gamma^2\sqrt{p_x p_y p_z}}\frac{ \left[\prod_i\int_{1}^{2p_i/\Delta'}  e^{-\frac{\Delta'}{p_i}(A_i-\frac{p_i}{\Delta'})^2}\,dA_i\right]\,\tilde{H}(\{p_i\},\{c_i\};\{A_i\})}{\prod_i\int_{1}^{2p_i/\Delta'}  e^{-\frac{\Delta'}{p_i}(A_i-\frac{p_i}{\Delta'})^2}\,dA_i}\,.\label{Bianchi_density_contapprox}
\ee

\subsubsection*{Expansion scalar}
From the very definition \eqref{Expansion} and the Hamilton's eqs. \eqref{BianchiHeqq}, the expansion scalar reads
\be
\theta=\frac{1}{\sqrt{p_1p_2p_3}}\frac{d\sqrt{p_i p_2 p_3}}{dt} =-\frac{8\pi \gamma}{2}\sum_j \frac{1}{p_j}\frac{\partial \mathcal{C}}{\partial c_j}=-\frac{8\pi \gamma}{2}\sum_j \frac{1}{p_j}\frac{\partial H_{BI}}{\partial c_j}\,,\label{expansion2}
\ee
and for $H_{BI}=H$ we obtain the actual expression for the QRLG model:
 
\be
\theta=\frac{1}{2\gamma}\frac{\left[\prod_i\int_{1}^{2p_i/\Delta'}  e^{-\frac{\Delta'}{p_i}(A_i-\frac{p_i}{\Delta'})^2}\,dA_i\right]\,\sum_j\frac{1}{p_j}\frac{\partial\tilde{H}}{\partial c_j}(\{p_i\},\{c_i\};\{A_i\})}{\prod_i\int_{1}^{2p_i/\Delta'}  e^{-\frac{\Delta'}{p_i}(A_i-\frac{p_i}{\Delta'})^2}\,dA_i}
\ee
where 
\be
\frac{1}{p_j}\frac{\partial\tilde{H}}{\partial c_j}(\{p_i\},\{c_i\};\{A_i\})=\sum_{i,k} \sqrt{\frac{p_k}{p_ip_j}}\sqrt{\frac{A_j A_i}{A_k}}\cos\left(c_j\sqrt{\frac{A_j}{A_iA_k}}\right)\sin\left(c_k\sqrt{\frac{A_k}{A_iA_j}}\right)\quad  i\neq j\neq k\,,\label{deHdecdivisop}
\ee
e.g. the $j=1$ component is

\begin{align}
\frac{1}{p_1}\frac{\partial\tilde{H}}{\partial c_1}&=\sqrt{\frac{p_3}{p_1p_2}}\sqrt{\frac{A_1A_2}{A_3}} \cos\left(c_1\sqrt{\frac{A_1}{A_2A_3}}\right)\sin\left(c_3\sqrt{\frac{A_3}{A_1A_2}}\right) \nonumber\\
&+\sqrt{\frac{p_2}{p_1p_3}}\sqrt{\frac{A_1A_3}{A_2}} \cos\left(c_1\sqrt{\frac{A_1}{A_2A_3}}\right)\sin\left(c_2\sqrt{\frac{A_2}{A_1A_3}}\right)\nonumber\,.
\end{align}

\subsubsection*{Shear}
Finally, using \eqref{BianchiHeqq} and \eqref{Shear}, we find the following expression for the shear:
\be
\sigma^2=\frac{(8\pi\gamma)^2}{3}\left[\left(\frac{\partial H}{p_1\partial c_1}-\frac{\partial H}{p_2\partial c_2}\right)^2+\left(\frac{\partial H}{p_2\partial c_2}-\frac{\partial H}{p_3\partial c_3}\right)^2+\left(\frac{\partial H}{p_3\partial c_3}-\frac{\partial H}{p_1\partial c_1}\right)^2\right]\label{Bianchishear}\,.
\ee

%%%%%%%%%%%%%%%%%%%%%%%%%%%%%%%%%%%%%%%%%%%%%%%%%%%%%%%%%%%%%%%%%%%%%%%%%%%%%%%%%%%%%%%%%%%%%%%%%%%%%%%%%%%%%%%%%%%%%%%%%%%%%

\section{Effective dynamics: numerical study}\label{Sec_effectivedynamics_numerical}

Here we address the effective dynamics of our model \eqref{constraint_Bianchi_QRLG}. In order to understand the choice of possible initial conditions, we first briefly review the Bianchi I dynamics in GR.

\subsection{Initial conditions and Kasner indices}

As it is well known (e.g. see \cite{Montani:2011zz}), when the classical Bianchi I geometry is sourced by a massless scalar field, the possible initial conditions for its associate Cauchy problem divide into the sets (of the starting points) of ``Kasner-like" and ``Kasner-unlike" solutions. Below we briefly review them following the notation of \cite{Chiou:2007mg}.

A straightforward computation reveals the vanishing of the following Poisson brackets:
\begin{equation}
\{p_{\phi},\mathcal{C}_{gr}\}\,,\qquad\{p_i c_i,\mathcal{C}_{gr}\}\qquad\forall\, i \,,
\end{equation} 
to which we associate the four constants of motion $p_{\phi},\, p_ic_i\,,$ which can be parametrized as
\begin{equation}
p_{\phi}:=\sqrt{8\pi}\mathcal{K}_{\phi}\,,\qquad p_ic_i:=8\pi\gamma\mathcal{K}_i\label{constantofmotion}
\end{equation}
where $\mathcal{K}_{\phi}:=k k_{\phi}\,,\mathcal{K}_i:=k k_i$ and $k$ a constant such that 
\begin{equation}
k_1+k_2+k_3=\pm 1\,.\label{kasnercondition1}
\end{equation}
The four real numbers $\{k_{\phi},k_i\}$ are called Kasner indices and in terms of them the vanishing of $\mathcal{C}_{gr}$ reads 
\begin{equation}
k_{\phi}^2+k_1^2+k_2^2+k_3^2=1\,.\label{kasnercondition2}
\end{equation}
Without loss of generality, we can stick to the case $\sum_i k_i =+1$ and $k_{\phi}>0$.  Kasner indices divide into two sets: the one where one $k_i$ is negative and the other two are positive, which is called ``Kasner-like", and the one where $k_i>0\, \forall\, i\,$, called ``Kasner-unlike" (note that this set includes also the isotropic case $k_i=1/3\,, i=1,2,3$). Classically, for both sets the singularity is not avoided as it is clear from the general solution \eqref{classicalsolution}, which we give below for the scale factors (from which $p_i(\phi)$ and $c_i(\phi)$ can be immediately obtained using \eqref{Bianchitriads} and \eqref{connectionGR}):
\begin{equation}
a_i(\phi)=a_i(\phi_0)\,e^{\sqrt{8\pi}\frac{k_i}{k_{\phi}}(\phi-\phi_0)}\,.\label{classicalsolution}
\end{equation}
In order to compare the dynamics provided by QRLG with the LQC one, we choose the same set of initial conditions for the Cauchy problem associated to the Hamiltons eqs. \eqref{BianchiHeqq}. This is a first order differential problem which admits a unique solution once seven initial conditions\footnote{For which we use a shorthand notation, e.g. $c_i(0):=c_i(\phi_0)=c_i(t_0)$.} $\{p_i(0),c_i(0),p_{\phi}\}$ are chosen such that $\mathcal{C}(p_i(0),c_i(0),p_{\phi})=0\,$. To be sure that this common set fulfills (approximately) both the LQC and QRLG constraints \eqref{constraintLQC},\eqref{constraint_Bianchi_QRLG}, we choose a set associated to a \textit{classical} universe, i.e. $p_i>>1, c_i<<1$ ($\mu_i c_i<<1$) where we know that the two constraints match. In this regime also GR holds and the possible initial conditions are those we were referring before, i.e. isotropic, Kasner-like and Kasner-unlike. For all these sets we follow the same strategy: we choose the same values of $\{p_i(0),c_i(0)\}$ for both models and obtain $p_{\phi\, lqc},\,p_{\phi\, qrlg}$ by imposing $\mathcal{C}_{lqc}(0)=0$ and $\mathcal{C}_{qrlg}(0)=0$, respectively.

\subsection{The isotropic case (ISO)}

\begin{figure}[H] 
	\begin{center}
		\begin{minipage}{0.4\textwidth}
			\centering
			\vspace{0.1cm}
			\includegraphics[width=8.5cm]{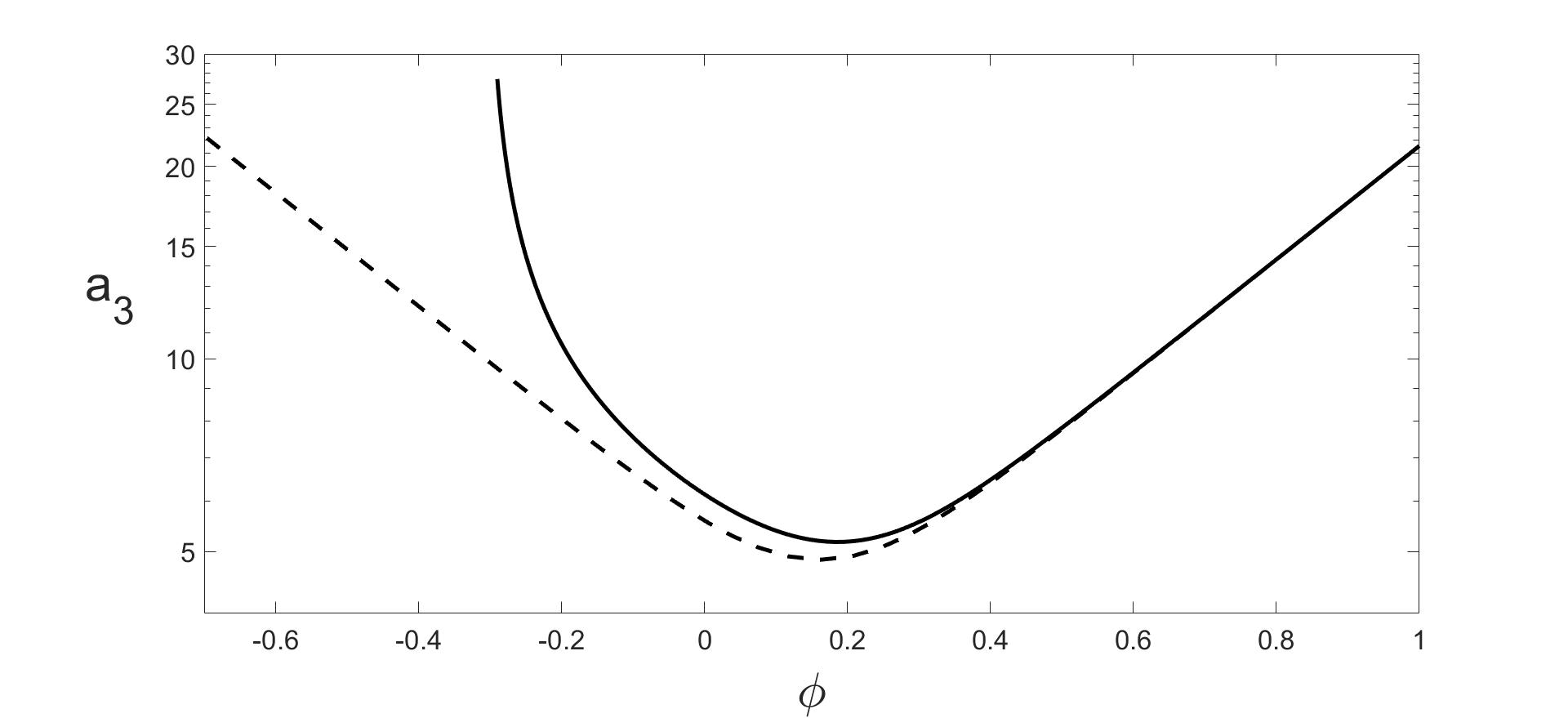}
		\end{minipage}
		\hspace{0.99cm} 
		\begin{minipage}{0.4 \textwidth}
			\centering
			\includegraphics[width=8.5cm]{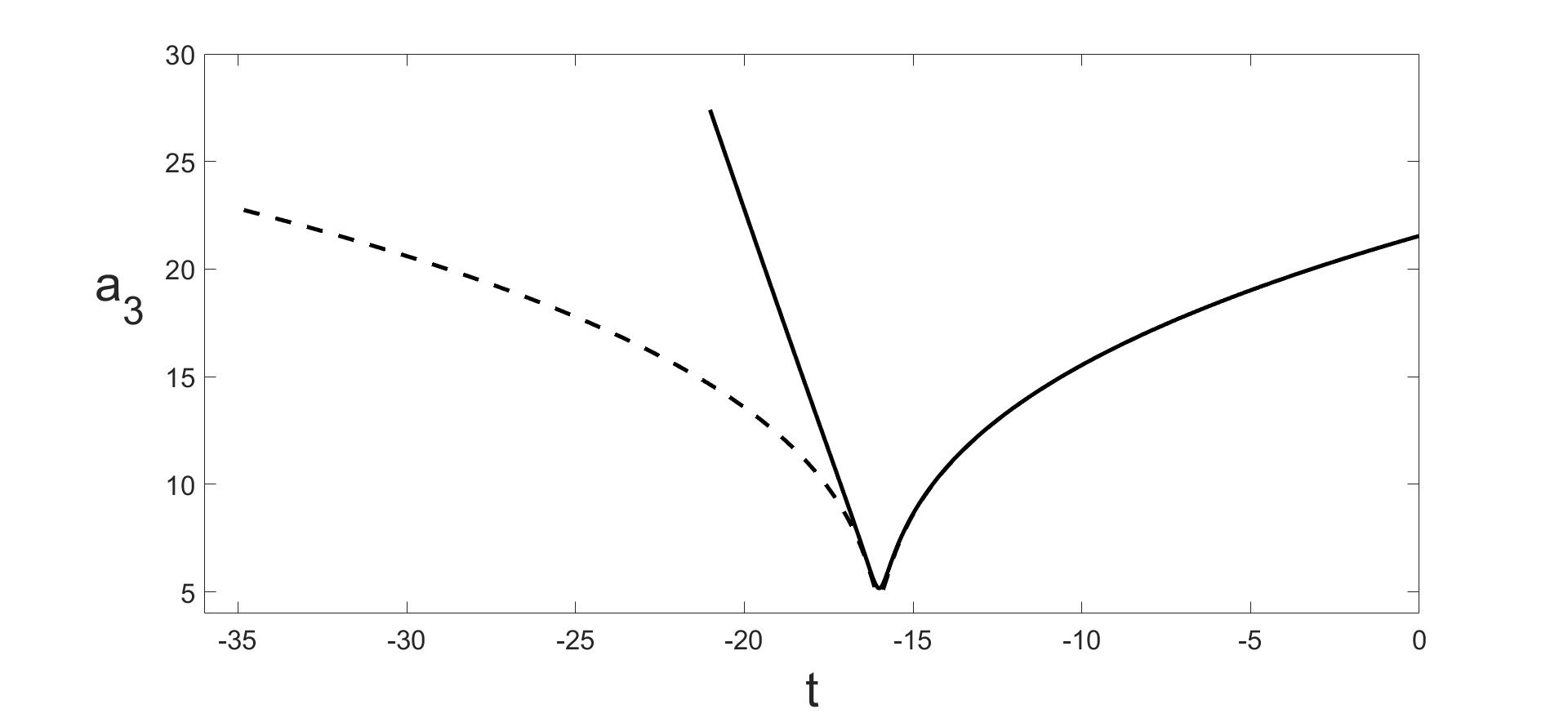}
		\end{minipage}
	\end{center}
	\caption{\small{\emph{Comparison between the QRLG dynamics (solid line) and the one provided by the LQC (dashed line) for isotropic initial conditions. Left panel: $a_3$ vs. the relational time $\phi\,$. Right panel: $a_3$ vs. the cosmological time $t$. In contrast to the LQC case, the QRLG evolution after the bounce is not longer Kasner un-like.}}}\label{fig_a3_iso}
\end{figure}

\begin{figure} [H]
	\centering
	\includegraphics[width=10cm]{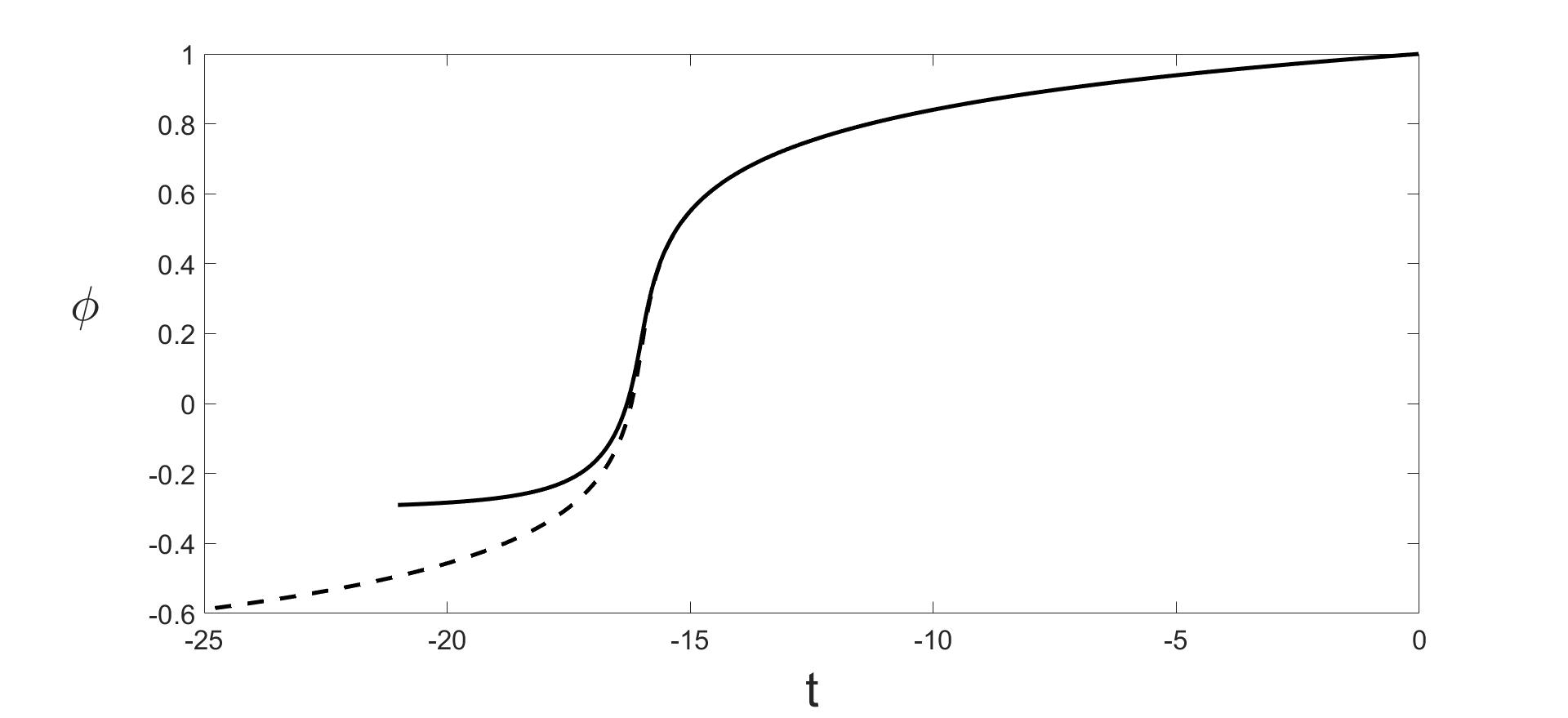}
	\caption{\small{\emph{Evolution of the field $\phi$ vs. $t$ for the ISO case. }}}\label{Fig_phiiso}
\end{figure}

\begin{figure}[H] 
 	\begin{center}
 		\begin{minipage}{0.4\textwidth}
 			\centering
 			\vspace{0.1cm}
 			\includegraphics[width=8.5cm]{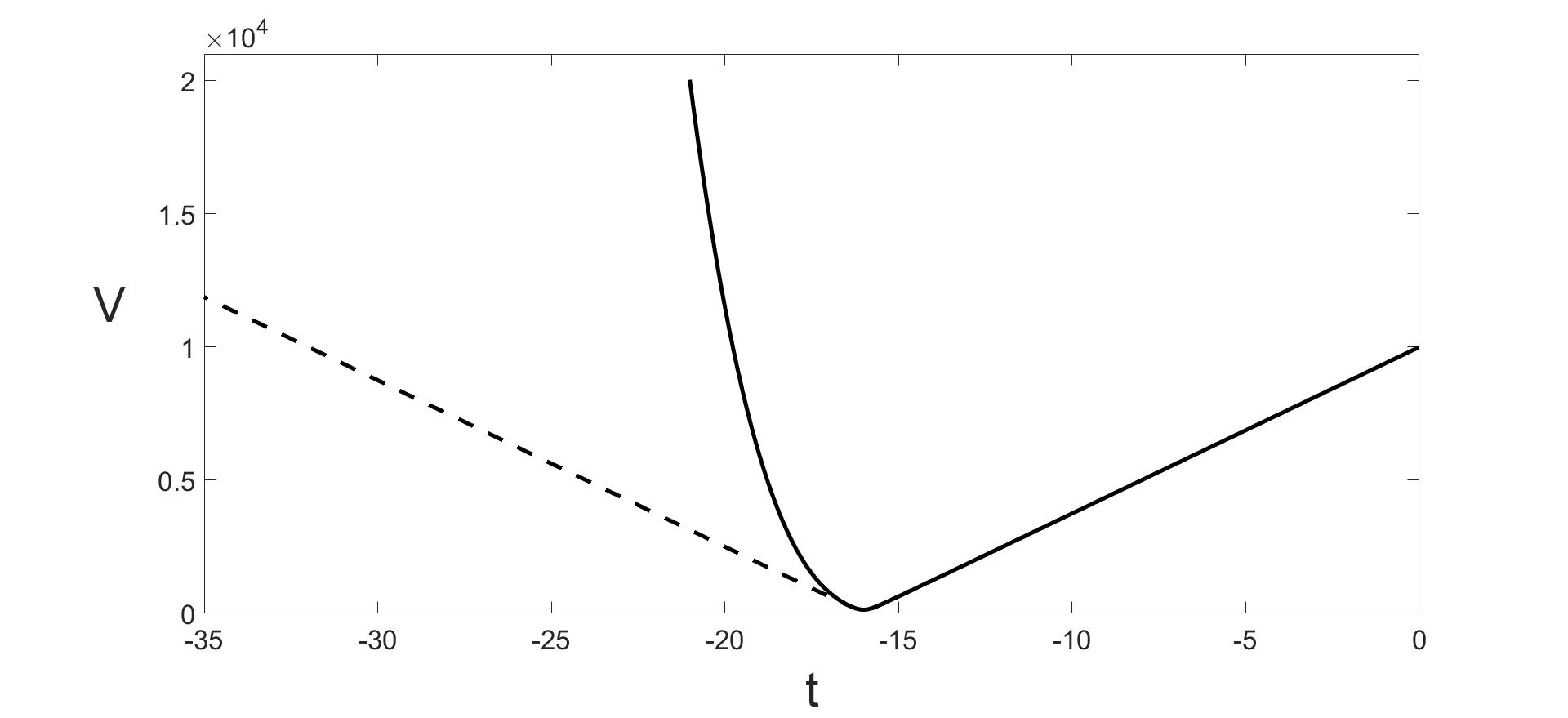}
 		\end{minipage}
 		\hspace{0.99cm} 
 		\begin{minipage}{0.4 \textwidth}
 			\centering
 			\includegraphics[width=8.5cm]{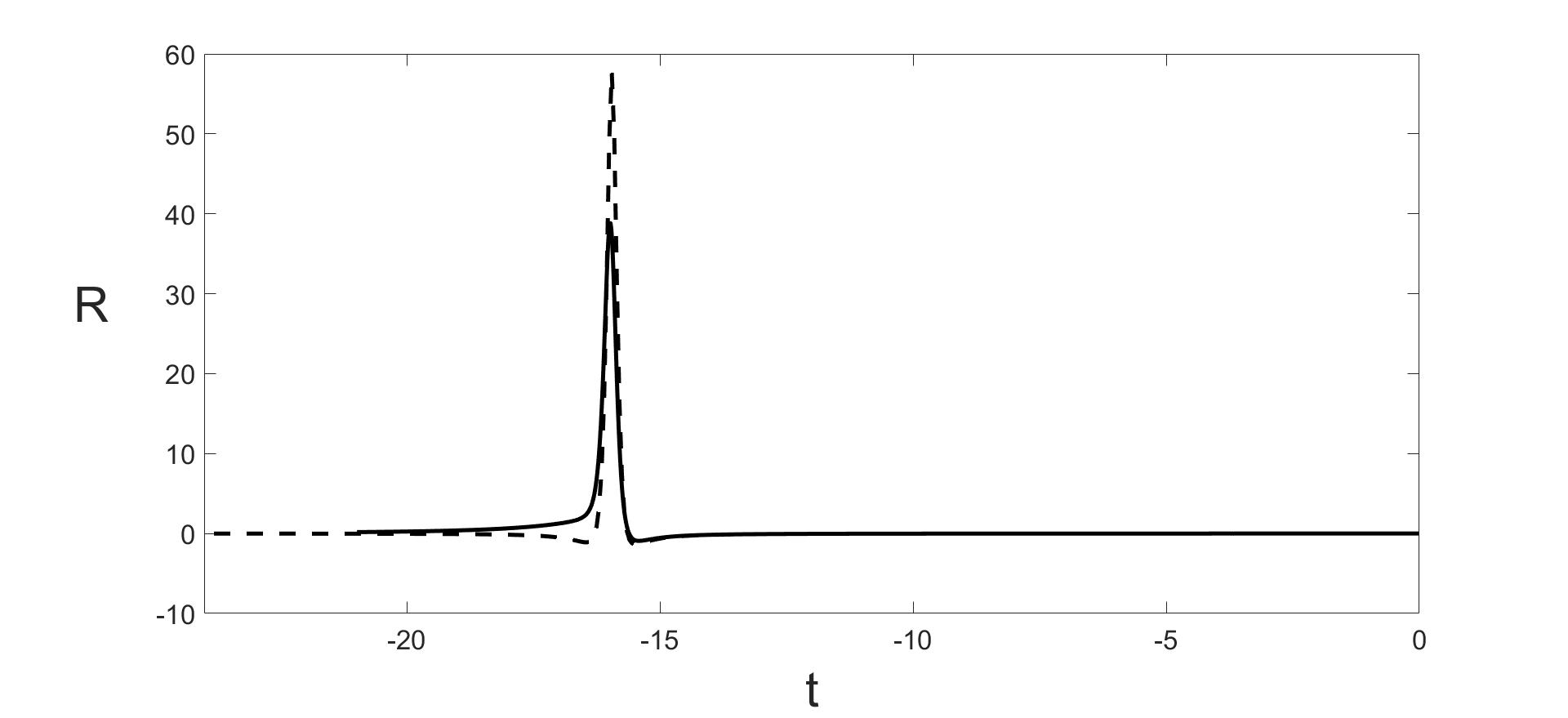}
 		\end{minipage}
 	\end{center}
 	\caption{\small{\emph{Comparison between the QRLG (solid line) and the LQC (dashed line) dynamics for isotropic initial conditions. Left panel: the volume $V$ vs. the cosmological time $t$. Right panel: the scalar curvature $R$ vs. $t$. Similar behaviours are found for the KUL and KL cases.}}}\label{fig_V_and_R_iso}
\end{figure}

 The chosen initial conditions for the isotropic case are: $p_i(0)=10^{8/3}\,,$ $c_i(0)=5\cdot10^{-5/3}\,,$ $\,p_{\phi \,qrlg}=101.78960\,,$ and $p_{\phi \,lqc}=101.78998\,.$ In fig.\ref{fig_a3_iso} we show the dynamics of the scale factor $a_3$, as it evolves in the relational time $\phi$ and in the cosmological time $t$ (the evolution along the other directions is exactly the same). After the bounce (which occurs approximately at the same time $t_B=-16$, $\phi_B=0.2$) the QRLG-Bianchi evolution shows a significant departure from the LQC one. In particular, looking at the left panel of fig.\ref{fig_a3_iso}, we see the two evolutions start differing from each other already a bit before the bounce.
  For the QRLG model, $\phi(t)$ approaches a constant value after the bounce time, i.e. for $t<-16$ (see fig.\ref{Fig_phiiso}). This explains why the $p_i(\phi)\,$ ($a_i(\phi)$) have a faster evolution than the $p_i(t)\,$ ($a_i(t)$), indeed, as the field reaches the plateau the dynamics ``accumulates" around $\phi\approx -0.3\,$. In particular, the scale factor grows linearly in time $t$ (with slope $-4.491$) after the bounce, explaining the vanishing behaviour of the scalar curvature $R\,$, 
 \begin{equation}
 R=2\left(H_1H_2+H_2H_3+H_3H_1+\sum_i \frac{\ddot{a}_i}{a_i}\right)\,,\label{R}
 \end{equation}
in the far past (see the right panel of fig.\ref{fig_V_and_R_iso}). As already observed in \cite{Chiou:2007mg}, the LQC evolution bridges two classical Bianchi I solutions, as it is clear from the dashed trajectories after and before the bounce in fig.\ref{fig_a3_iso}. This is not the case for the QRLG model, which is very peculiar. Going backwards in time, the QRLG Bianchi I universe starts as an initial classical Bianchi I and after the bounce undergoes an expansion that it is neither KL nor KUL, i.e. it never approaches the classical solution \eqref{classicalsolution} for any Kasner index set. As we will see, this turns out to be a general feature of the QRLG dynamics, observed also in the evolutions associated to KUL and KL initial conditions (see fig.\ref{fig_p_a_cpos} and fig.\ref{fig_p_a_cneg}).

\subsection{The Kasner-unlike (KUL) case}

\begin{figure}[H] 
	\begin{center}
		\begin{minipage}{0.4\textwidth}
			\centering
			\vspace{0.1cm}
			\includegraphics[width=8.5cm]{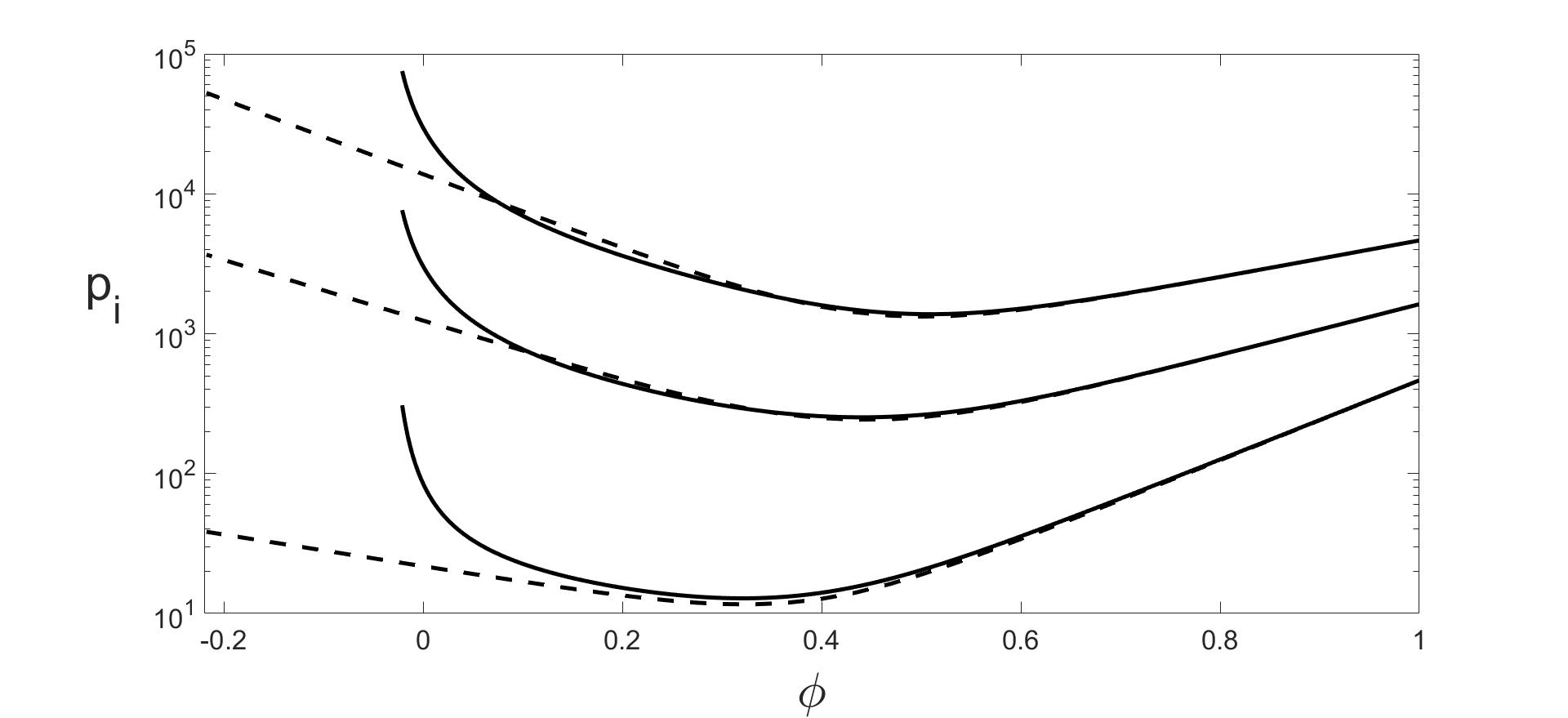}
		\end{minipage}
		\hspace{0.99cm} 
		\begin{minipage}{0.4 \textwidth}
			\centering
			\includegraphics[width=8.5cm]{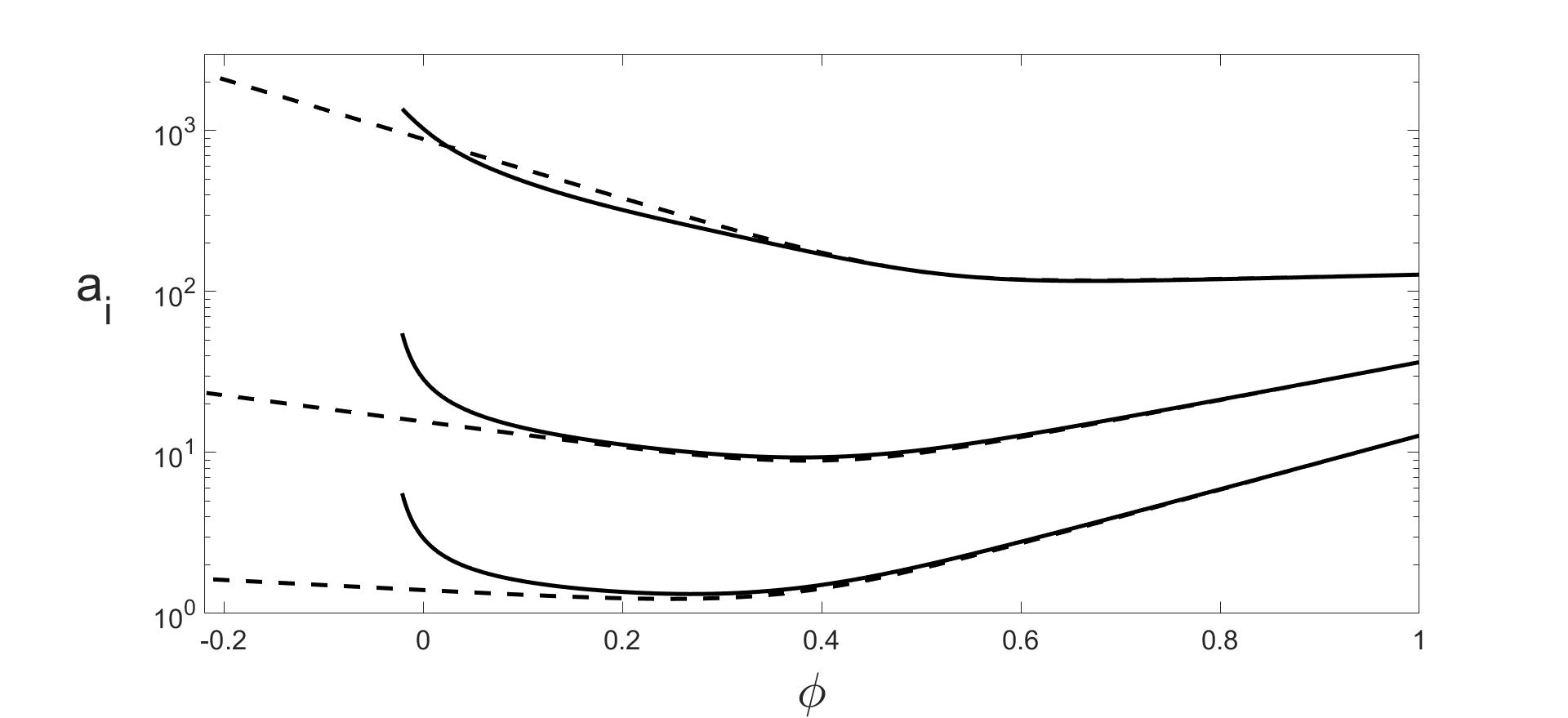}
		\end{minipage}
	\end{center}
	\caption{\small{\emph{Comparison between the QRLG (solid line) and the LQC (dashed line) dynamics for Kasner un-like initial conditions. Left panel from top to bottom: $p_3,p_2,p_1$ vs. the relational time $\phi$. Right panel from bottom to top: the scale factors $a_3,a_2,a_1$ vs. the relational time $\phi$. The Kasner indices for the LQC evolution are: $k_1(0)=0.0467\,,$ $k_2(0)=0.3925\,,$ $k_3(0)=0.5608\,$ and $k_1=-0.6203\,,$ $k_2=-0.2741\,,$ $k_3=-0.1056\,.$}}}\label{fig_p_a_cpos}
\end{figure}

\begin{figure}[H] 
	\begin{center}
		\begin{minipage}{0.4\textwidth}
			\centering
			\vspace{0.1cm}
			\includegraphics[width=8.5cm]{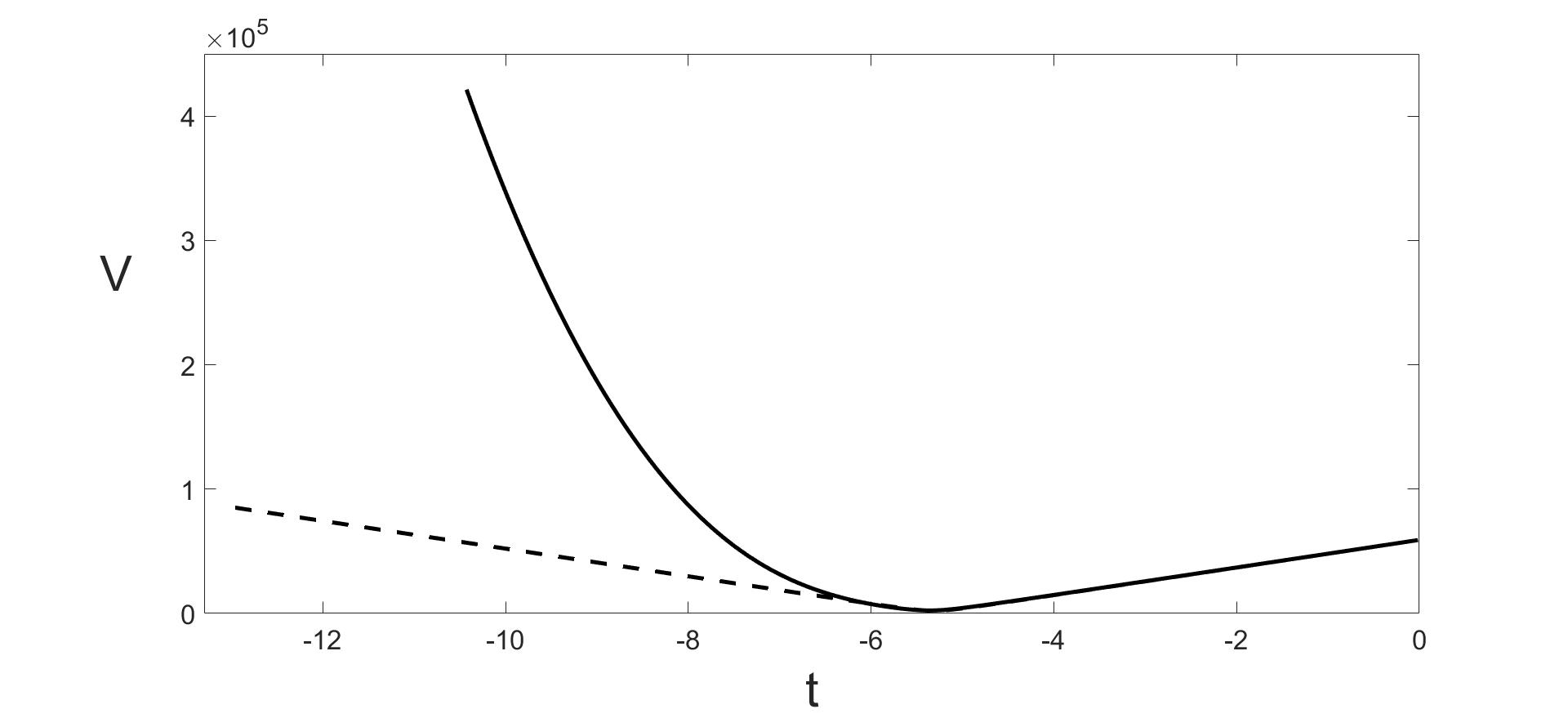}
		\end{minipage}
		\hspace{0.99cm} 
		\begin{minipage}{0.4 \textwidth}
			\centering
			\includegraphics[width=8.5cm]{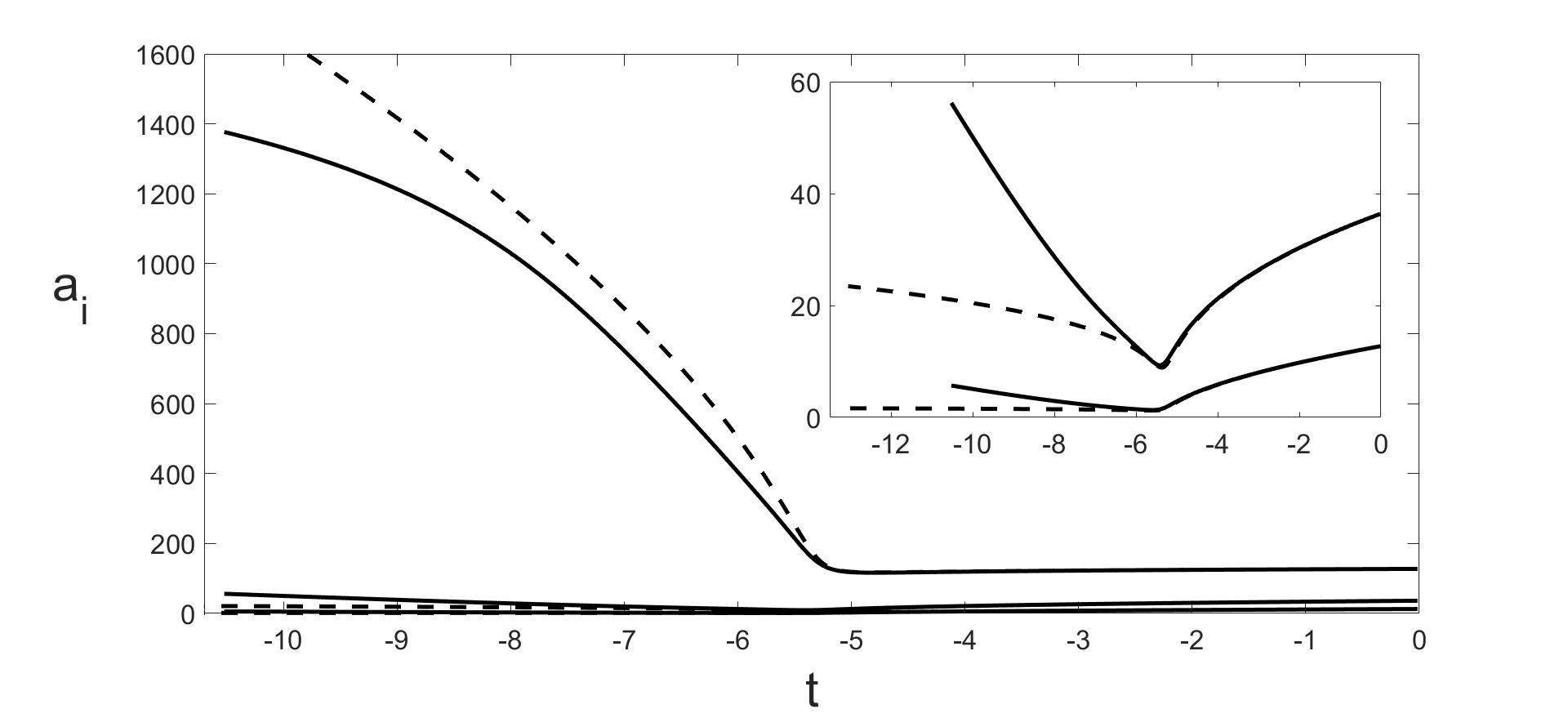}
		\end{minipage}
	\end{center}
	\caption{\small{\emph{Comparison between the QRLG (solid line) and the LQC (dashed line) dynamics for Kasner un-like initial conditions. Left panel: $V$ vs. $t$. Right panel from top to bottom: scale factors $a_1,a_2,a_3$ vs. the cosmological time $t$. The inset depicts the evolution of $a_2$ and $a_3$ vs. $t$.}}}\label{fig_V_and_a_t_cpos}
\end{figure}

Here we show the dynamics associated to the following anisotropic set of initial conditions:  $p_{1}(0)= 
10^{8/3}\,,$ $p_2(0)=3.5\cdot 10^{8/3}\,,$ $p_3(0)=10\cdot10^{8/3}\,,$ $c_1(0)=12.5\cdot10^{-5/3}\,,$ $c_2(0)=30\cdot10^{-5/3}\,,$ $c_3(0)=15\cdot10^{-5/3}\,,$ $p_{\phi\, lqc}=1616.8821 \,,$ and $p_{\phi\, qrlg}=1616.7799\,.$ In fig.\ref{fig_p_a_cpos} the physical areas\footnote{for which unitary fiducial leghts $\tilde{L}_i$ have been chosen.} $p_i$ and the scale factors $a_i$ are plotted in the left and right panel, respectively. We see that the LQC model joins two classical Bianchi I universes associated to (initial) Kasner indices $k_i(0)$ and (final) Kasner indices $k_i\,$, such that\footnote{confirming what already observed in \cite{Chiou:2007mg}.} $k_i=k_i(0)-2/3$. Instead, the QRLG evolution starts as a classical Bianchi I with $k_i(0)$ like LQC but departs from the latter after the bounce and \textit{accelerates} (going backwards in the relational time $\phi$). In fig.\ref{fig_V_and_a_t_cpos} the volume and the scale factors are plotted as they evolve in the cosmological time $t$, where the evolutions are power laws.

\subsection{The Kasner-like (KL) case}

\begin{figure}[H] 
	\begin{center}
		\begin{minipage}{0.4\textwidth}
			\centering
			\vspace{0.1cm}
			\includegraphics[width=8cm]{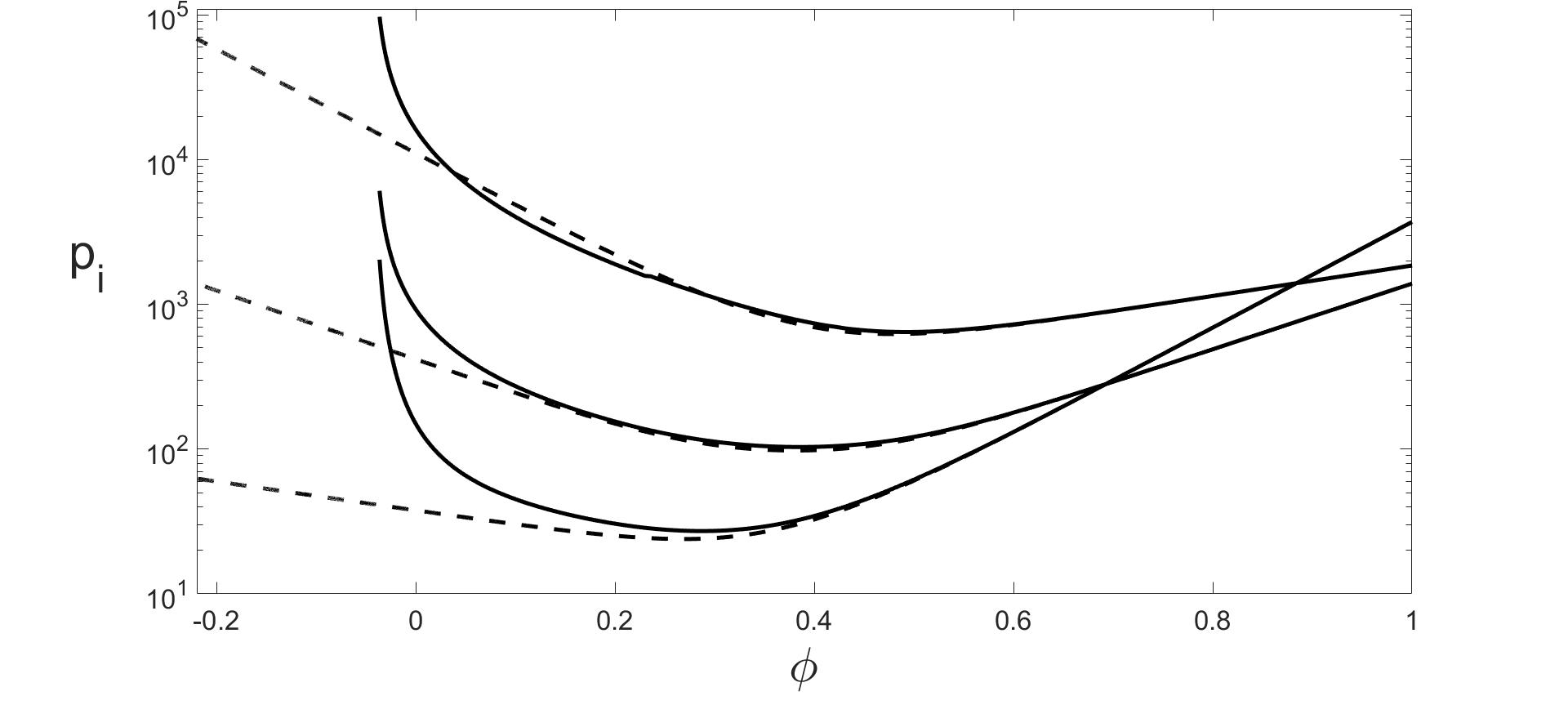}
		\end{minipage}
		\hspace{0.99cm} 
		\begin{minipage}{0.4 \textwidth}
			\centering
			\includegraphics[width=8.5cm]{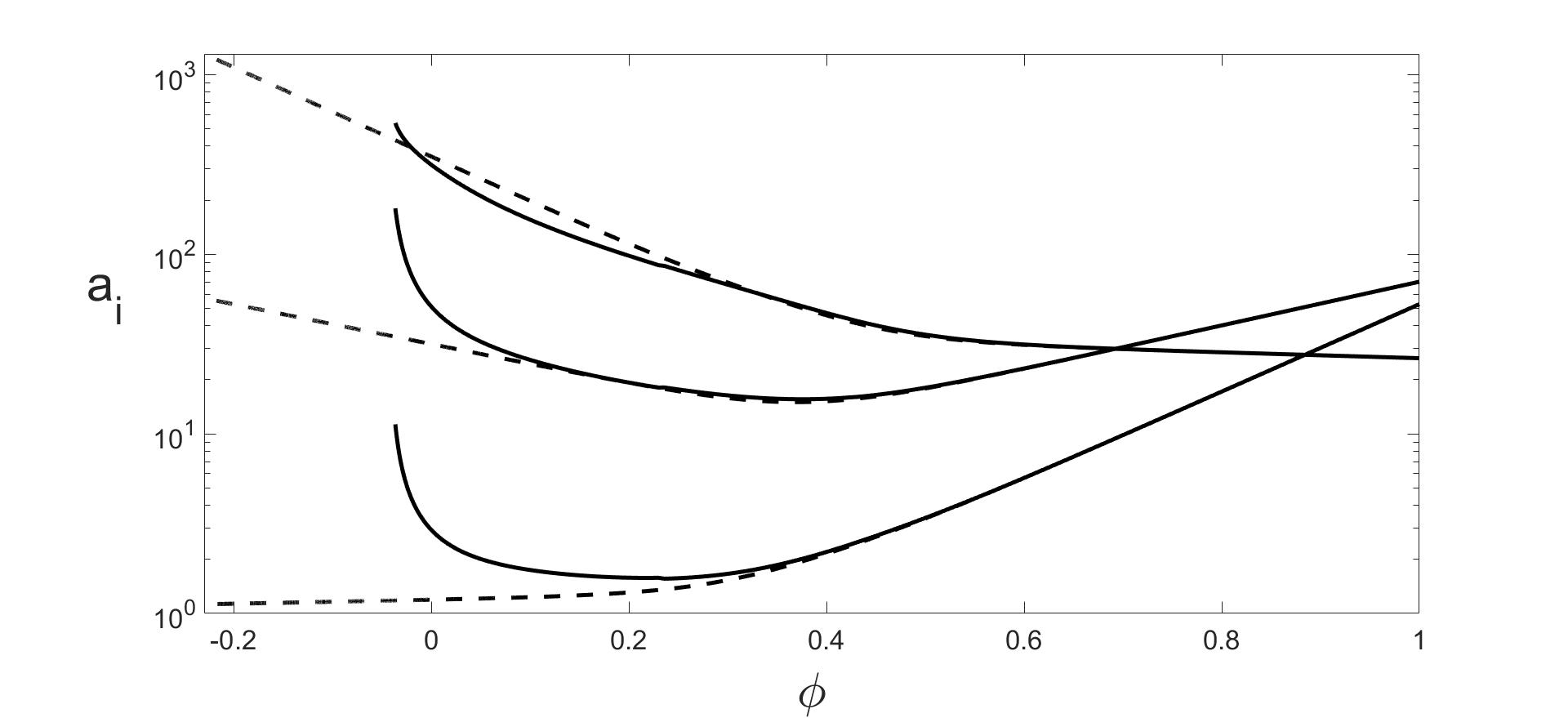}
		\end{minipage}
	\end{center}
	\caption{\small{\emph{Comparison between the dynamics of QRLG (solid line) and LQC (dashed line) for Kasner-like initial conditions. Left panel from top to bottom: $p_1,p_3,p_2$ vs. the relational time $\phi$. Right panel from top to bottom: evolution of the scale factors $a_2,a_3,a_1$ vs. the relational time $\phi$. The Kasner indices for the LQC evolution are: $k_1(0)=0.6978\,,$ $k_2(0)=-0.0465\,,$ $k_3(0)=0.3488\,$ and $k_1=0.0311\,,$ $k_2=-0.7133\,,$ $k_3=-0.3178\,.$ Note that the scale factors in the QRLG model are non vanishing in \textit{all} directions, contrarily to what happens in the LQC case where $a_1$ goes to zero.}}}\label{fig_p_a_cneg}
\end{figure}

Finally, we present the anisotropic case associated to an initially contracting direction, e.g. to a negative $c_2(0)$. The chosen initial conditions for this case are $p_{1}(0)=4\cdot10^{8/3}\,,$ $p_{2}(0)=8\cdot10^{8/3}\,,$ $p_{3}(0)=3\cdot10^{8/3}\,,$ $c_1(0)=30\cdot10^{-5/3}\,,$ $c_2(0)=-10^{-5/3}\,,$ $c_3(0)=20\cdot10^{-5/3}\,,$ and $p_{\phi\, qrlg}=891.9694\,,$ $p_{\phi\,lqc}=891.98209\,.$ Even though the evolutions of the areas $p_i$ (showed in the left panel of fig.\ref{fig_p_a_cneg}) are similar to those of the KUL case, the scale factors (plotted in the right panel of the same figure) reveal a peculiar feature: each directional scale factor is non-vanishing, contrary to what we observe for the LQC model where $a_1$ goes to zero after its bounce (see the bottom dashed line in the right panel).

\begin{figure}[H] 
	\begin{center}
		\begin{minipage}{0.4\textwidth}
			\centering
			\vspace{0.1cm}
			\includegraphics[width=8.5cm]{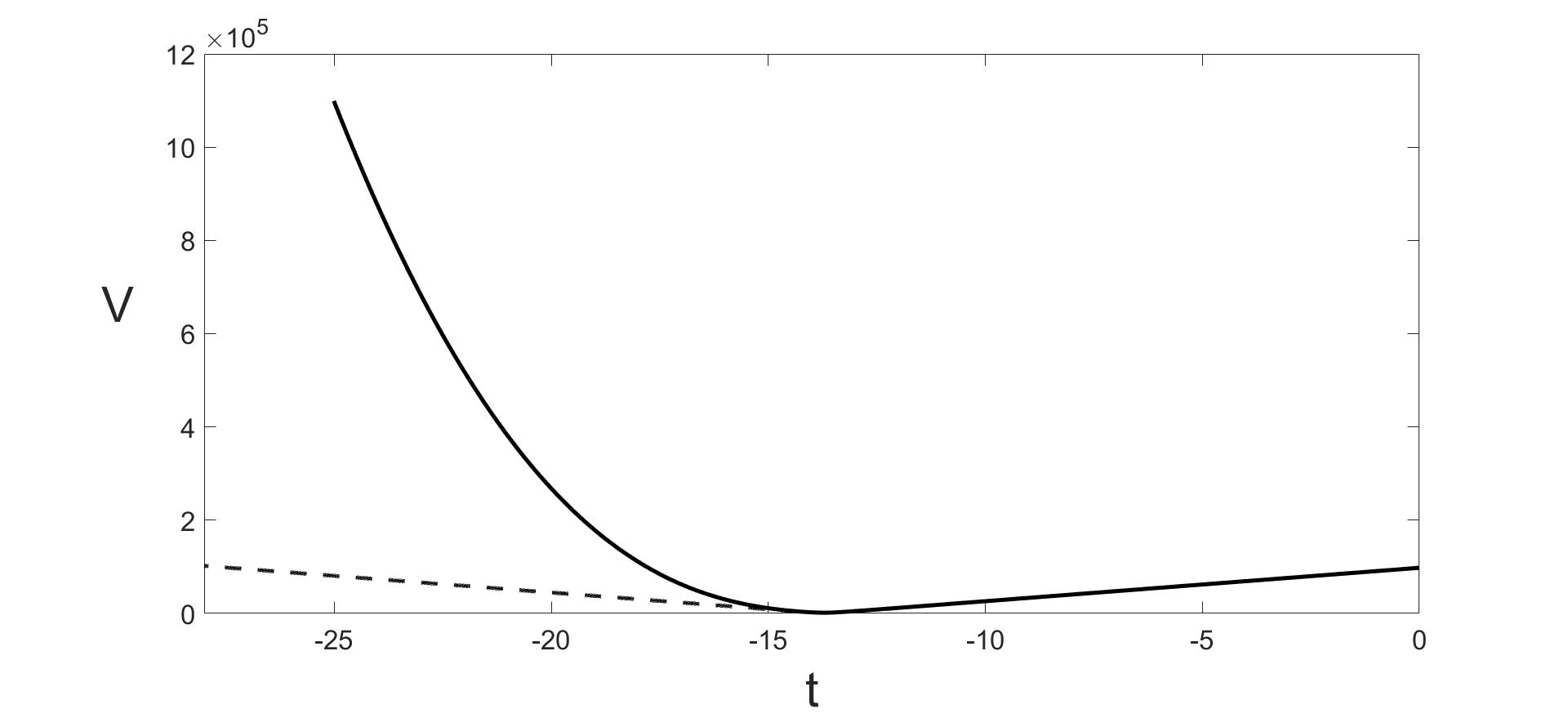}
		\end{minipage}
		\hspace{0.99cm} 
		\begin{minipage}{0.4 \textwidth}
			\centering
			\includegraphics[width=8.5cm]{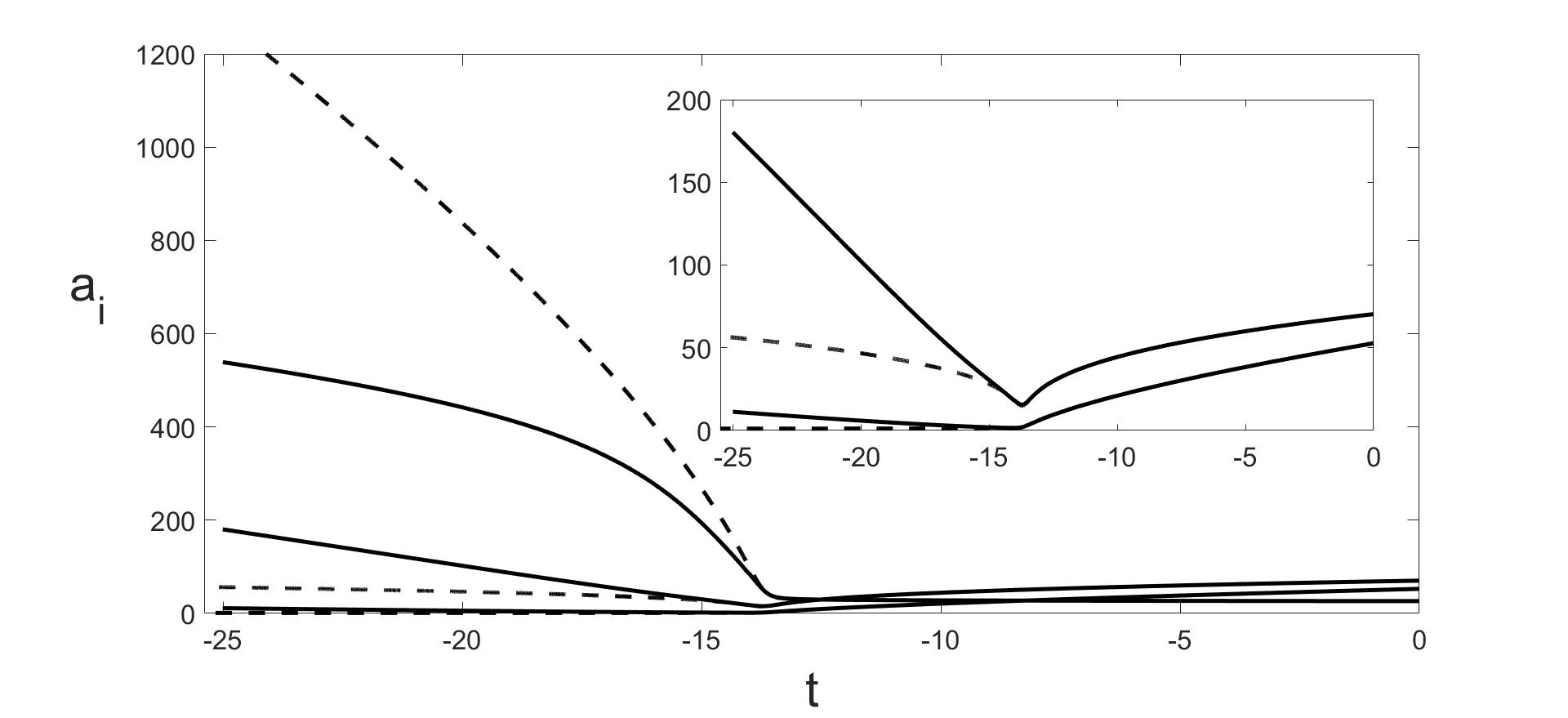}
		\end{minipage}
	\end{center}
	\caption{\small{\emph{Comparison between the dynamics of QRLG (solid line) and LQC (dashed line) for Kasner like initial conditions. Left panel: $V$ vs. $t$. Right panel from top to bottom: the evolution of the scale factors $a_2,a_3,a_1$ vs. $t$. The inset depicts the evolution of $a_3$ and $a_1$ vs. $t$.}}}\label{fig_V_and_a_t_cneg}
\end{figure}

\begin{figure}[H] 
	\begin{center}
		\begin{minipage}{0.4\textwidth}
			\centering
			\vspace{0.1cm}
			\includegraphics[width=8.5cm]{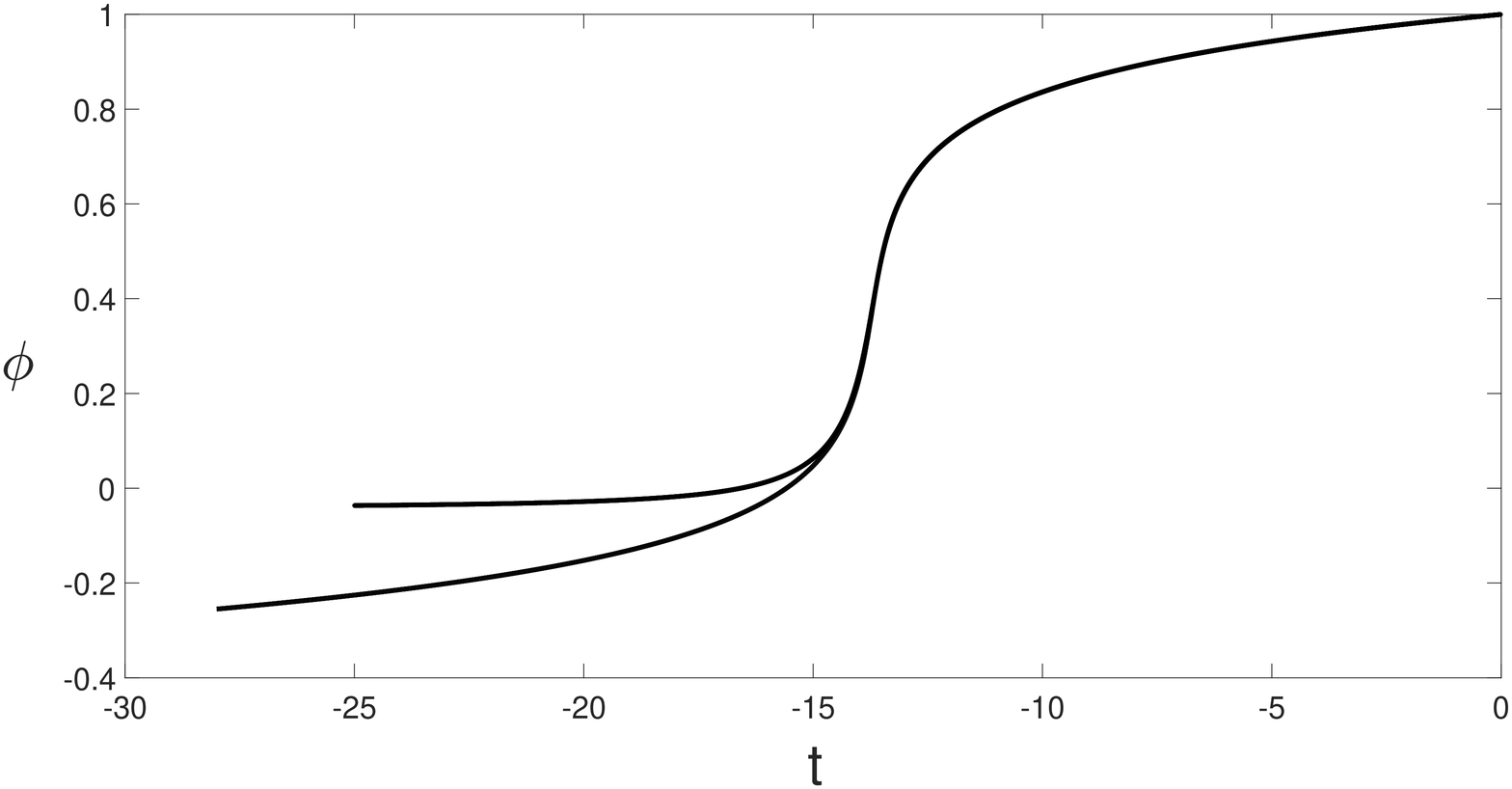}
		\end{minipage}
		\hspace{0.99cm} 
		\begin{minipage}{0.4 \textwidth}
			\centering
			\includegraphics[width=8.5cm]{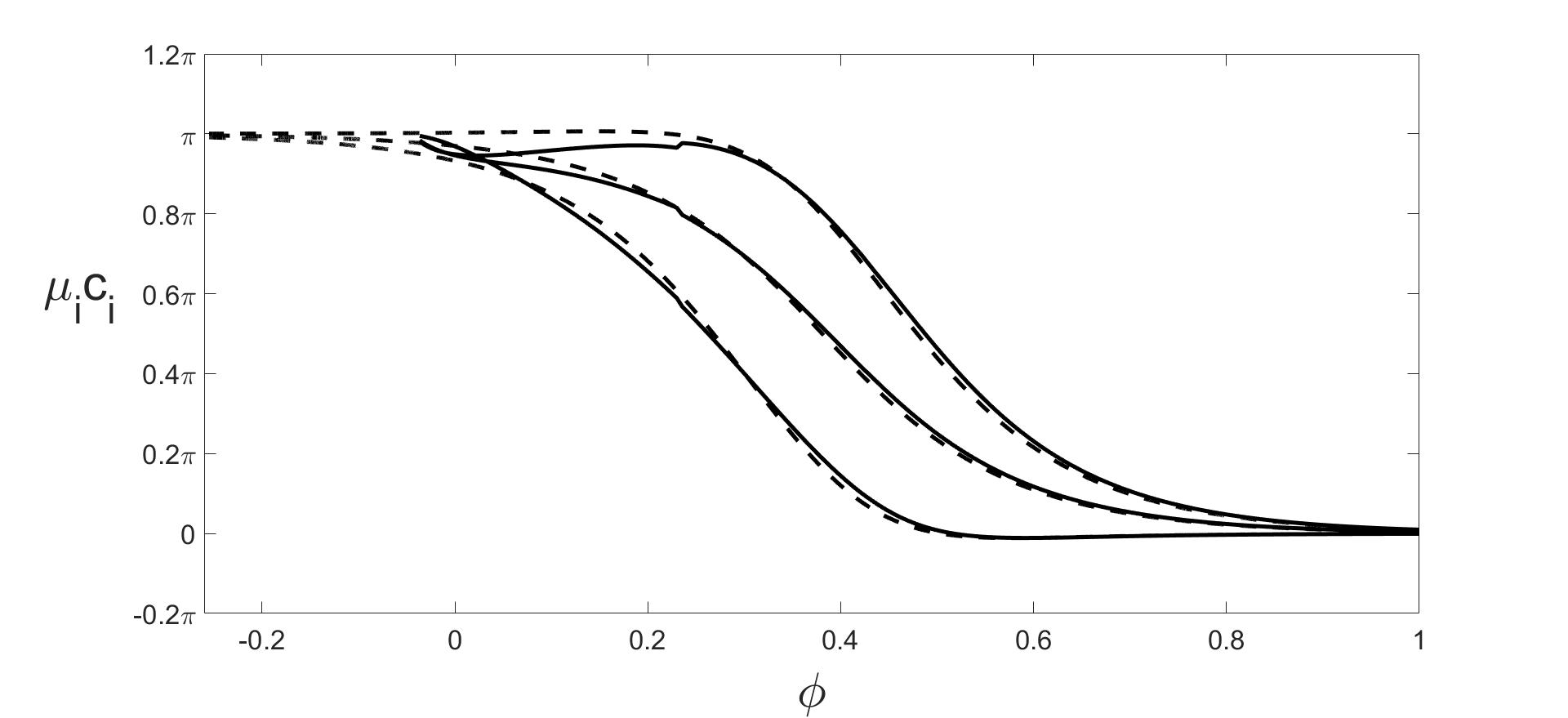}
		\end{minipage}
	\end{center}
	\caption{\small{\emph{Comparison between the dynamics of QRLG (solid line) and LQC (dashed line) for Kasner like initial conditions. Left panel: evolution of the field $\phi$ vs. the cosmological time $t$. Right panel from top to bottom: evolution of the semiclassical parameters $\mu_1 c_1,\mu_3c_3,\mu_2 c_2$ vs. $t$, where $\mu:=\bar{\mu},\mu'$. Similar behaviours are obtained for the ISO and KUL cases.}}}\label{fig_phi_and_mui_cneg}
\end{figure}

In closing, a natural question arises: why after the bounces the QRLG model does not follow a semiclassical Bianchi I evolution if its volume goes back to macroscopic values? A look at its Hamiltonian \eqref{HBianchi_QRLG_contapprox} is enough for the answer. Indeed, the QRLG Hamiltonian is an integral function whose next-to-the leading order contribution for large volumes provide a first order correction to the Bianchi I LQC Hamiltonian that is proportional to the semiclassical parameter $\mu'_ic_i$, where $\mu'_i:=\sqrt{\Delta'p_i/(p_jp_k)}\,,i\neq j\neq k$. A straighforward computation reveals \cite{Alesci:2017kzc} that the ratio of this correction over the magnitude of a $\sin^2$-term in the LQC Hamiltonian \eqref{constraintLQC} goes like $V^{-2/3}\mu'_ic_i$. Thus, for finite volumes $V$, both models match only when $\mu'_ic_i\rightarrow 0\,$. In the right panel of fig.\ref{fig_phi_and_mui_cneg} we clearly see that this is no longer true for $\phi<<\phi_{B}\approx 0.3\,$. For macroscopic times, the LQC evolution matches the GR one. The closer we are to $\phi\approx \pi\,$ the better the LQC Hamiltonian is approximated by the classical one \eqref{constraintGR}, but the QRLG model does not because its infinite contributions coming from the integral \eqref{HBianchi_QRLG_contapprox} are $\mathcal{O}(1)\,$.

\subsection{Technical details}

Two strategies have been followed. On the one hand, we numerically solved the system of eqs.\eqref{BianchiHeqq} by means of a fourth order Runge-Kutta Merson method in order to obtain $p_{2}$, $p_{3}$, $c_{1}$, $c_{2}$ and $c_{3}$, both for the QRLG and LQC model. In the first case, $H_{qrlg}$ of \eqref{Bianchiconstraint} is given by eq.\eqref{HBianchi_QRLG_contapprox}, while in the second case, $H_{lqc}$ is provided by eq.\eqref{constraintLQC}. On the other hand, we have obtained $p_{1}$ directly from \eqref{Bianchiconstraint}, which is regarded as an integral equation for $p_{1}$, by means of the Trust-Region Dogleg method implemented in the function \texttt{fsolve} of MATLAB\textsuperscript{®} \cite{Powell:1970jth}. For the QRLG case, we evaluated the integrals by using the functions \texttt{integral1},\texttt{integral2} and \texttt{integral3} encoded in MATLAB\textsuperscript{®}, which make use of a global adaptive quadrature rule based on a Gauss-Kronrod scheme. We retained the default values for the absolute the relative tolerances, being $1e^{-10}$ and $1e^{-6}$ respectively. For the QRLG case, we have choosen $\Delta$$t=0.0158$ for all the three cases analyzed, i.e. the ISO, KL, KUL case. As far as LQC concerns, we selected $\Delta$$t=0.05$ for both the ISO and KUL case, and $\Delta$$t=0.0016$ for the KL one, where the complete system of eqs. \eqref{BianchiHeqq} have been solved to obtain $p_{1}$ as well. The choice of the time step and the strategy of the solution are dictated by the necessity of keeping the $C_{qrlg}$ of the order of at least $10^{-8}$, trying to minimize the computational time. The initial conditions of the simulations, together with the values of $p_{\phi}$ used for QRLG and LQC are listed in table \ref{Table_ic}.

\begin{table}[H]
	\begin{center}
		\begin{tabular}{|p{0.8cm}|p{1.5cm}|p{0.8cm}|p{1.5cm}|p{0.8cm}|p{1.5cm}|}
			\hline
			\multicolumn{6}{|c|}{Initial conditions} \\
			\hline
			\multicolumn{2}{|c|}{ISO case} & \multicolumn{2}{|c|}{KUL case} & \multicolumn{2}{|c|}{KL case}\\
			\hline
			$p_{1}(0)$ & $\alpha^{2/3}$ & $p_{1}(0)$ & $\alpha^{2/3}$ & $p_{1}(0)$ & $4\, \alpha^{2/3}$\\ \hline
			$p_{2}(0)$ & $\alpha^{2/3}$ & $p_{2}(0)$ & $3.5\,\alpha^{2/3}$ & $p_{2}(0)$ & $8\, \alpha^{2/3}$\\ \hline
			$p_{3}(0)$ & $\alpha^{2/3}$ & $p_{3}(0)$ & $10\, \alpha^{2/3}$ & $p_{3}(0)$ & $3\, \alpha^{2/3}$\\ \hline
			$c_{1}(0)$ & $\beta \alpha^{1/3}$ & $c_{1}(0)$ & $2.5\, \beta \alpha^{1/3}$ & $c_{1}(0)$ & $6\, \beta \alpha^{1/3}$\\ \hline
			$c_{2}(0)$ & $\beta \alpha^{1/3}$ & $c_{2}(0)$ & $6\, \beta \alpha^{1/3}$ & $c_{2}(0)$ & $-0.2\, \beta \alpha^{1/3}$\\ \hline
			$c_{3}(0)$ & $\beta \alpha^{1/3}$ & $c_{3}(0)$ & $3\, \beta \alpha^{1/3}$ & $c_{3}(0)$ & $4\, \beta \alpha^{1/3}$\\ \hline
			$p_{\phi \,qrlg}$ & $101.78960$ & $p_{\phi\, qrlg}$ & $1616.7799 $ & $p_{\phi\, qrlg}$ & $891.96940$\\ \hline
			$p_{\phi \,lqc}$ & $101.78998$ & $p_{\phi\, lqc}$ & $1616.8821$ & $p_{\phi \,lqc}$ & $891.98209$\\
			\hline
		\end{tabular}	
		\caption{\small{\emph{Initial conditions for the three cases discussed in the main text. $\alpha:=10^{4}\,,$ $\beta:=5\cdot10^{-3}$. }}} \label{Table_ic}
	\end{center}	
\end{table}

\begin{figure}[H] 
	\begin{center}
		\begin{minipage}{0.4\textwidth}
			\centering
			\vspace{0.1cm}
			\includegraphics[width=8.5cm]{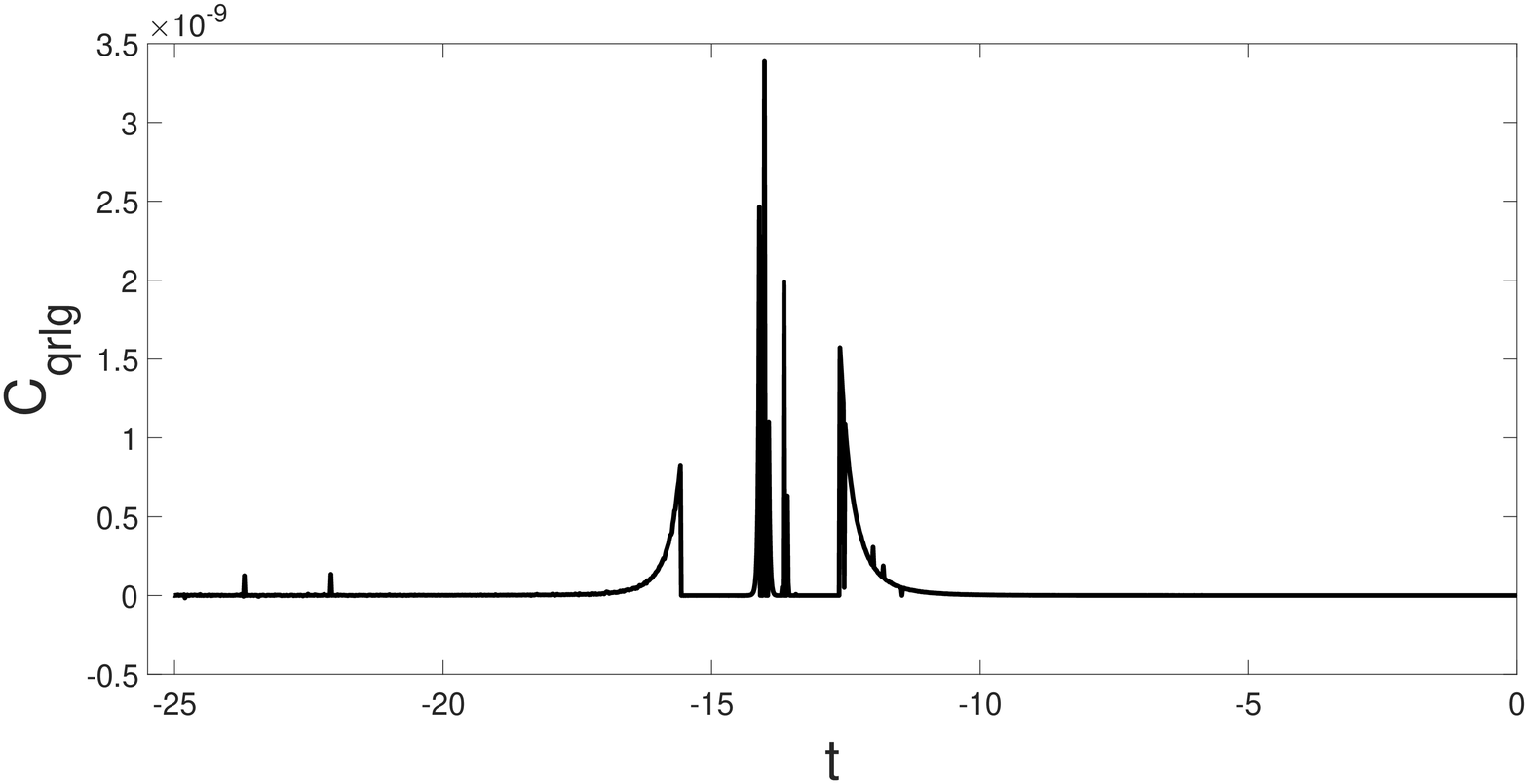}
		\end{minipage}
		\hspace{0.99cm} 
		\begin{minipage}{0.4 \textwidth}
			\centering
			\includegraphics[width=8.5cm]{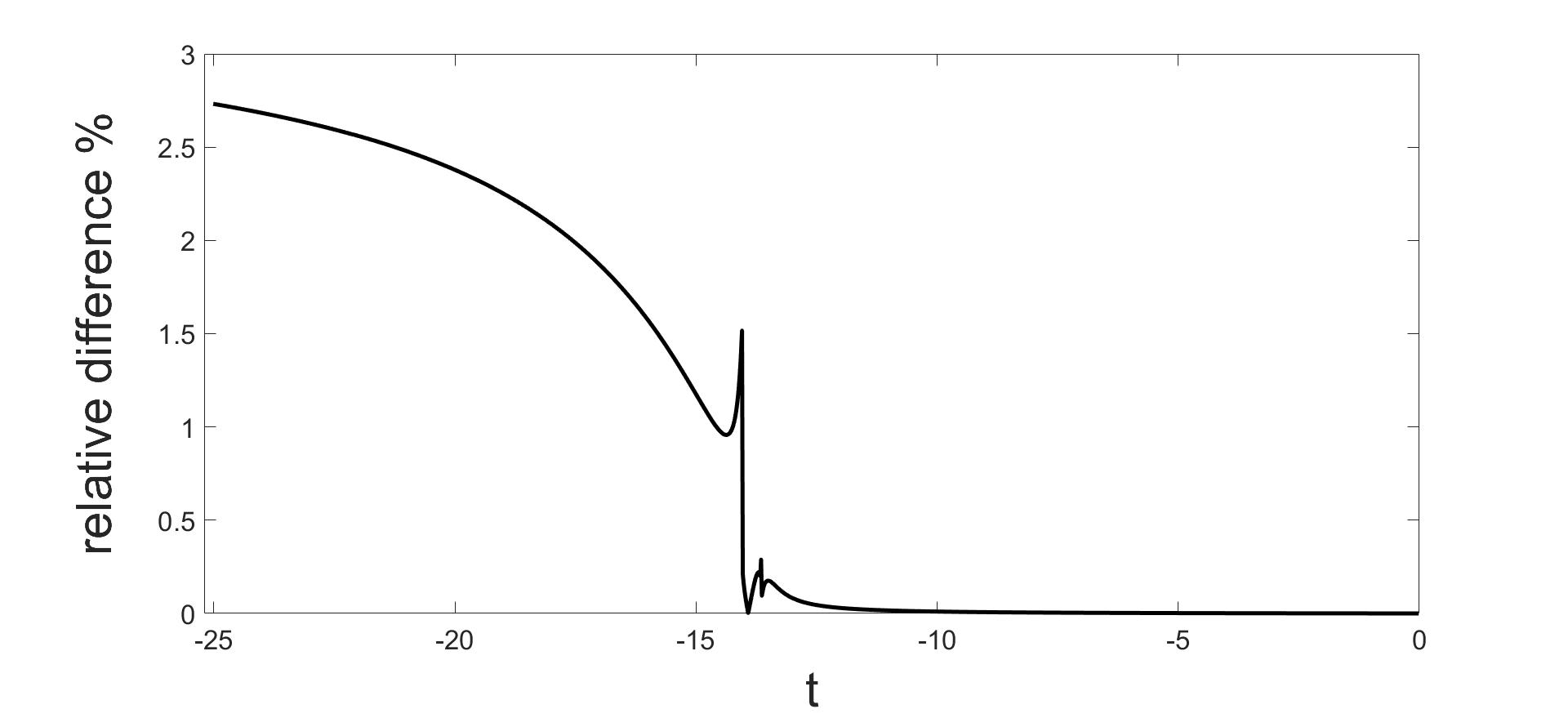}
		\end{minipage}
	\end{center}
	\caption{\small{\emph{Left panel: evolution of the $C_{qrlg}$ vs. $t$ where $p_{1}$ has been obtained from eq. \eqref{constraint_Bianchi_QRLG}.  Right panel: percentage of the relative difference of the volume obtained with the two methods vs. $t$. In one case $p_{1}$ is computed from eq. \eqref{constraint_Bianchi_QRLG} while in the other case $p_{1}$ is computed from eq. \eqref{BianchiHeqq} together with $p_{2}$, $p_{3}$, $c_{1}$, $c_{2}$ and $c_{3}$ by means of the fourth order Runge-Kutta-Merson method.}}}\label{fig_costraint_volume}
\end{figure}	

In figure \ref{fig_costraint_volume} we show the evolution of the $C_{qrlg}$ vs. $t$ for the KL case. It can be seen that the values of $C_{qrlg}$ are very low, precisely of the order of $10^{-9}$ or less. However, we want to remark that the way we solved the dynamical problem always delivered small values of $C_{qrlg}$, even if hypothetical errors were present in our code. This happens because $p_{1}$ is computed directly from eq. \eqref{constraint_Bianchi_QRLG} . 
Therefore, we compared the numerical solution obtained with this strategy with the one given by the solution of the complete system of eqs. \eqref{BianchiHeqq}, which does not depend on the \eqref{constraint_Bianchi_QRLG}. In case of  errors in the code, $C_{qrlg}$ would strongly deviate from $0$ with this procedure. Therefore, a comparison between the numerical solutions of the governing equations obtained with the two strategies immediately indicates whether the solution computed with the first method is correct, since both solutions must match. Figure \ref{fig_costraint_volume} depicts the percentage of the relative difference of the volume computed with both strategies. It can be seen that the numerical results match well during the whole period of the evolution, with a maximum relative difference of approximately $2.7\%$. In addition, the numerical solution obtained by calculating $p_{1}$ directly from \eqref{constraint_Bianchi_QRLG} gives $C_{qrlg}$ of the order of $10^{-9}$ or less. Instead, the solution obtained by solving the complete system of equations \eqref{BianchiHeqq} delivered $C_{qrlg}$ of the order of $10^{-4}$ with the chosen time step and tolerances of the integrals. By utilizing this strategy for the QRLG case, the order of $C_{qrlg}$ does not decrease significantly when the time step $\Delta$$t$ is reduced. Thus, we have choosen to solve $p_{1}$ from \eqref{constraint_Bianchi_QRLG} in all the simulations for all the QRLG cases, even if the computational time required was higher, since the Trust-Region Dogleg method is an iterative technique. This fact is more severe for the QRLG case, since numerical simulations required several days on a common workstation, while in the LQC case the computational time was just of the order of seconds. 

%In the QRLG case, the numerical evaluation of integrals significantly increases the computational time compared to the LQC one.

% To conclude, we numerically solved the system of eqs.\eqref{BianchiHeqq} for the vacuum case in order to compute $p_{1}$, $p_{2}$, $p_{3}$, $c_{1}$, $c_{2}$ and $c_{3}$, both for cases QRLG and LQC. This resulted in a significant reduction of time, since the convergence of the functions \texttt{integral1}, \texttt{integral2} and \texttt{integral3} and of the Trust-Region Dogleg method is extremely slow. Nevertheless, we obtained maximum values of $C_{qrlg}$ and $C_{lqc}$ of the order of $10^{-5}$ and $10^{-6},$ respectively, by utilizing $\Delta t=0.0158$.   

%%%%%%%%%%%%%%%%%%%%%%%%%%%%%%%%%%%%%%%%%%%%%%%%%%%%%%%%%%%%%%%%%%%%%%%%%%%%%%%%%%%%%%%%%%%%%%%%%%%%%%%%%%%%%%%%%%%%%%%%%%%%%

\section{Analytical upper bounds for the QRLG-Bianchi I model}\label{Sec_Upperbounds}

Here we proove that energy density, expansion scalar and shear are bounded phase space functions along the effective dynamics provided by the QRLG constraint (\ref{HBianchi_QRLG_disc}). To begin with, let us introduce the following quantity (where $b:=c/\sqrt{p}$):
\be
H^{FLRW}_{disc,A} (A^{max}(V),b):=-\frac{3}{8\pi \gamma^2}\left(\frac{\Delta'A^{max}}{2}\right)^{1/2} \,\frac{\sum_1^{A^{max}}\left(\begin{matrix}A^{max} \\ A \end{matrix}\right)A \sin^2[b\left(\frac{\Delta'A^{max}}{2A}\right)^{1/2}]}{\sum_1^{A^{max}}\left(\begin{matrix}A^{max} \\ A \end{matrix}\right)}\,,\label{ACTUAL_disc_FullHarea}
\ee
which is the effective QRLG-Hamiltonian for the FLRW model, as provided by the (area counting) statistical regularization scheme \cite{Alesci:2017kzc}. In order to show the boundness of the energy density $\rho^{disc}$ for the QRLG-Bianchi I model, i.e. of \eqref{Bianchi_density_contapprox} with gaussians replaced by binomials as in \eqref{HBianchi_QRLG_disc}, we will use as preliminary lemma the boundness of the energy density associated to \eqref{ACTUAL_disc_FullHarea}, which we proove below.

\subsubsection*{Energy density upper bound for the QRLG-FLRW model}

The energy density for the FLRW model is given by the ratio of \eqref{ACTUAL_disc_FullHarea} over the volume $V=\left(\Delta'\frac{A^{max}}{2}\right)^{3/2}$ \cite{Alesci:2017kzc}, followed by a sign change, i.e.
\be
\rho^{FLRW}_{disc,A}=\frac{3}{8\pi \gamma^2}\left(\frac{\Delta'A^{max}}{2}\right)^{-1} \,\frac{1}{2^{A^{max}}-1}\sum_1^{A^{max}}\left(\begin{matrix}A^{max} \\ A \end{matrix}\right)A \sin^2[b\left(\frac{\Delta'A^{max}}{2A}\right)^{1/2}]
\ee
where we have used 
\be
\sum_{1}^{A^{max}}\left(\begin{matrix}A^{max} \\ A \end{matrix}\right)=2^{A^{max}}-1\,.\label{binomialesommato}
\ee
Clearly, the following holds:
\be
\rho^{FLRW}_{disc,A}\leq\frac{3}{8\pi \gamma^2}\left(\frac{\Delta'A^{max}}{2}\right)^{-1} \,\frac{1}{2^{A^{max}}-1}\sum_1^{A^{max}}\left(\begin{matrix}A^{max} \\ A \end{matrix}\right)A
\ee
and using
\be
\sum_1^{A^{max}}\left(\begin{matrix}A^{max} \\ A \end{matrix}\right)A=A^{max}\,2^{A^{max}-1}\,,
\ee
we find
\be
\rho^{FLRW}_{disc,A}\leq\frac{3}{4\Delta'\pi \gamma^2}\max_{A^{max}\geq 1} S(A^{max})\,,
\ee
having called
\be
S(A^{max}):=\frac{2^{A^{max}-1}}{2^{A^{max}}-1}\,,\label{sequence}
\ee
which is a decreasing monotonic sequence whose maximum value is $1$, reached at $A^{max}=1$. Thus, we end up with the following upper bound:
\be
\rho^{FLRW}_{disc,A}\leq\frac{3}{4\Delta'\pi \gamma^2} \,.
\ee

\subsection{Energy density upper bound for the QRLG-Bianchi I model}

For Bianchi I we use expression \eqref{HBianchi_QRLG_disc}, change its sign and divide it for the volume 
$$V=\sqrt{p_1p_2p_3}= \left(\frac{\Delta'}{2}\right)^{3/2} \sqrt{A^{max}_1A^{max}_2A^{max}_3}.$$
Proceeding analougously to the QRLG-FLRW case, we arrive at the following inequality for the QRLG-Bianchi I energy density:
\be
\rho\leq \frac{1}{4\Delta'\pi \gamma^2}\,\, 3\, \max_{A^{max}_1\geq 1} S(A^{max}_1)\,,
\ee
where $S$ is the sequence \eqref{sequence}. Thus, we find the same bound we had for the isotropic case, i.e.
\be
\rho\leq\frac{3}{4\Delta'\pi \gamma^2}=0.6874\,.\label{bound_rho}
\ee

\subsection{Expansion scalar and shear upper bounds}

\begin{figure}
  	\centering
 	\includegraphics[width=10cm]{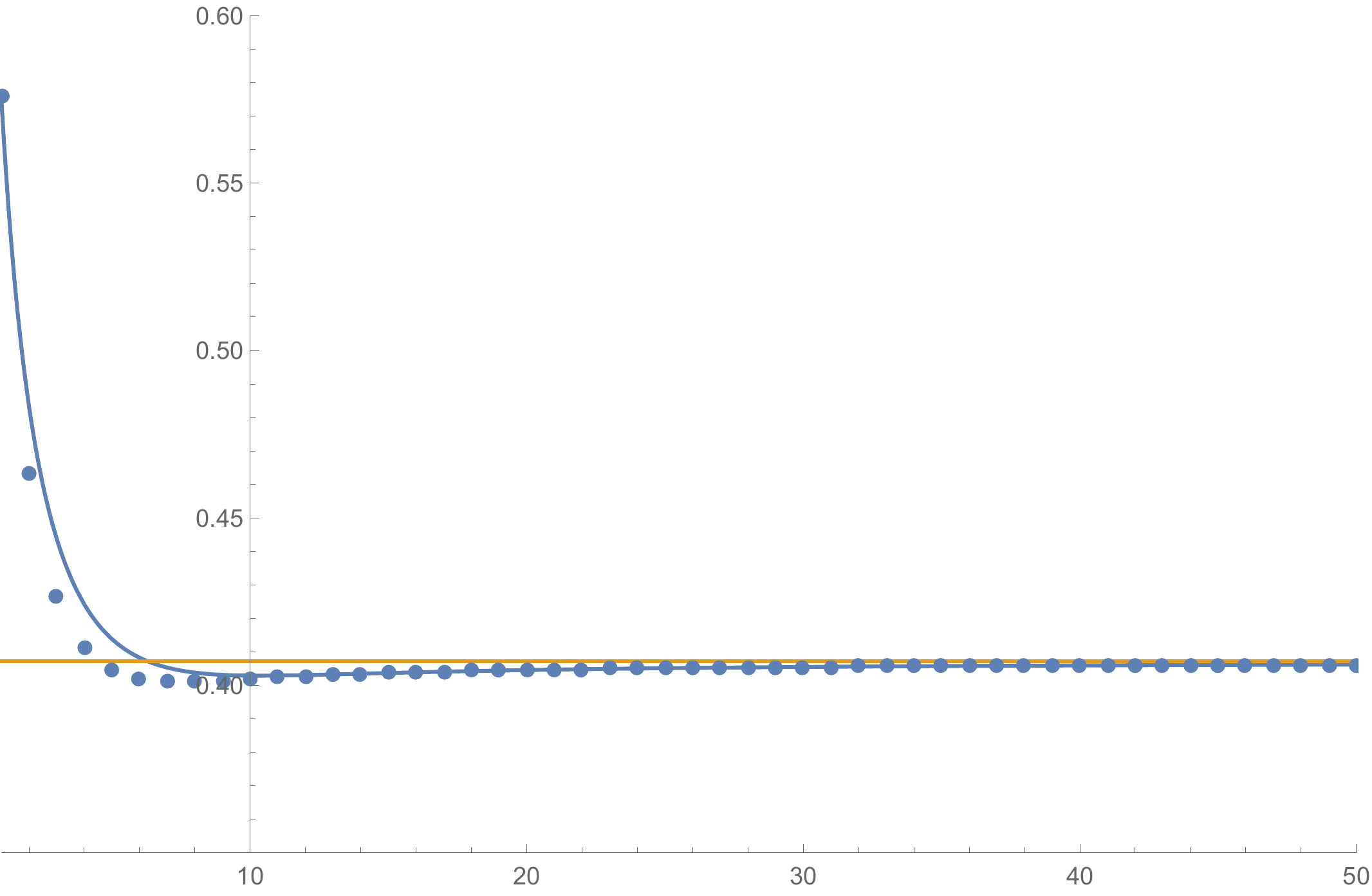}
 	\caption{\small{\emph{The sequence $I^{disc}_1(A^{max}_1)$ and its continuous approximation $I_1(A^{max}_1)$ defined in the text, together with their common asymptote $I_1=1/\sqrt{\Delta'}$. The maximum value of $I^{disc}_1$ is $0.576$, reached at $A^{max}_1=1$. Note the sequence starts matching its continuous approximation already for $A^{max}_1>10.$}}}\label{Fig_I1}
 \end{figure}

\begin{figure}
 	\centering
 \includegraphics[width=10cm]{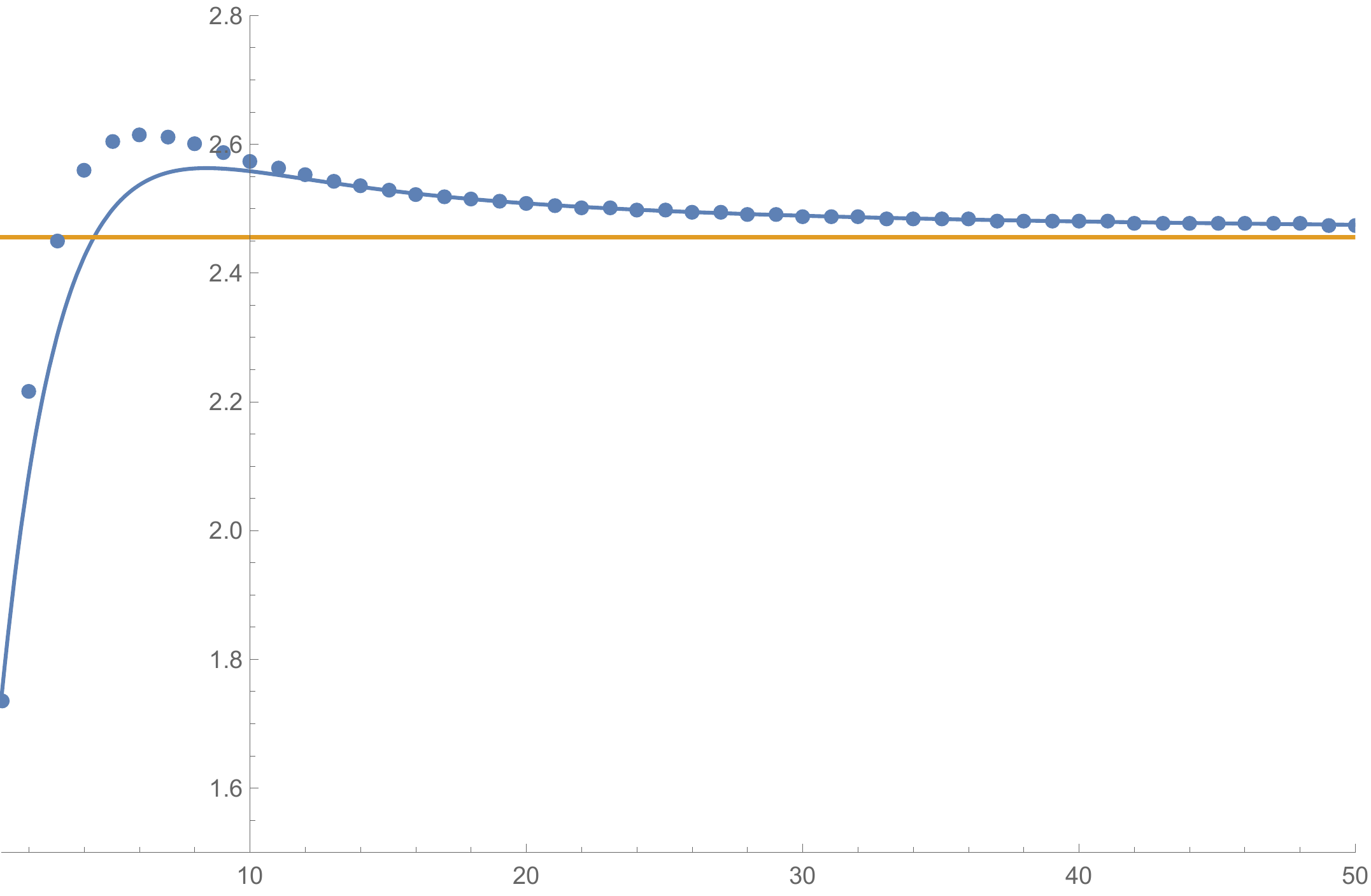}
 \caption{\small{\emph{The sequence $I^{disc}_2(A^{max}_2)$ and its continuous approximation $I_2(A^{max}_2)$ defined in the text, together with their common asymptote $I_2=\sqrt{\Delta'}$. The maximum value of $I^{disc}_2$ is $2.616$ reached at $A^{max}_2=6$ (the relative difference with the maximum of $I_2$ is only around $2\%$). Note the sequence starts matching its continuous approximation already for $A^{max}_2>12$.}}}\label{Fig_I2}
 \end{figure}

For the sake of clarity, here we start working within the continuous approximation. In order to show the boundness of $\theta$ and $\sigma^2$, it is enough to find an upper bound for the term $\frac{\partial H}{p_j\partial c_j}$, as it is clear from expressions \eqref{expansion2} and \eqref{Bianchishear}. From the definitions \eqref{HBianchi_QRLG_contapprox} and \eqref{deHdecdivisop}, it follows that
\be
\left|\frac{\partial H}{p_j\partial c_j}\right| \leq\frac{1}{8\pi\gamma^2}\sum_{i,k}\sqrt{\frac{p_k}{p_jp_i}}\frac{\left[\prod_i\int_{1}^{2p_i/\Delta'}  e^{-\frac{\Delta'}{p_i}(A_i-\frac{p_i}{\Delta'})^2}\,dA_i\right]\,\sqrt{\frac{A_jA_i}{A_k}}}{\prod_i\int_{1}^{2p_i/\Delta'}  e^{-\frac{\Delta'}{p_i}(A_i-\frac{p_i}{\Delta'})^2}\,dA_i}\quad i\neq j\neq k\,.
\ee
 For any given $j$, the r.h.s. is a sum of two terms which is symmetric under $k\leftrightarrow i$, e.g. for $j=1$
we have
$$\left|\frac{\partial H}{p_1\partial c_1}\right|\leq\frac{1}{8\pi\gamma^2}\left[I_1(p_1)I_2(p_2)I_1(p_3)+I_1(p_1)I_2(p_3)I_1(p_2) \right]$$ 
where
\begin{eqnarray}
I_{1}(p_1)&:=&\frac{1}{\sqrt{p_1}}\frac{\int_{1}^{2p_1/\Delta'}  e^{-\frac{\Delta'}{p_1}(A_1-\frac{p_1}{\Delta'})^2}\,\sqrt{A_1}\,dA_1}{\int_{1}^{2p_1/\Delta'}  e^{-\frac{\Delta'}{p_1}(A_1-\frac{p_1}{\Delta'})^2}\,dA_1}\,,\label{I1}\\
I_{2}(p_2)&:=&\sqrt{p_2}\frac{\int_{1}^{2p_2/\Delta'}  e^{-\frac{\Delta'}{p_2}(A_2-\frac{p_2}{\Delta'})^2}\,\frac{1}{\sqrt{A_2}}\,dA_2}{\int_{1}^{2p_2/\Delta'}  e^{-\frac{\Delta'}{p_2}(A_2-\frac{p_2}{\Delta'})^2}\,dA_2}\,.\label{I2}
\end{eqnarray}
 $I_1$ and $I_2$ are integral functions whose inspection at their boundaries is enough to understand whether they are bounded or not. Their asymptotic behaviour as $p_i\rightarrow\infty$ may be obtained with the Laplace method:
\be
\lim_{p_1\rightarrow \infty} I_1(p_1)=\frac{1}{\sqrt{\Delta'}}\,,\quad \lim_{p_2\rightarrow \infty} I_2(p_2)=\sqrt{\Delta'}\,.
\ee
Still, the limits at $p_i=\frac{\Delta'}{2}$ remain. At those points the continuous approximation \eqref{HBianchi_QRLG_contapprox} is no longer reliable (a priori) and the exact definition \eqref{HBianchi_QRLG_disc} must be taken into account, which implies the inspection of the discrete versions of functions $I_1$ and $I_2$, i.e.

\begin{eqnarray}
I_1^{disc}(A_1^{max})&:=&\frac{\sqrt{2}}{\sqrt{A^{max}_1\Delta'}(2^{A^{max}_1}-1)}\sum_{1}^{A^{max}_1}\left(\begin{matrix}A^{max}_1 \\ A_1 \end{matrix}\right)\sqrt{A_1}\,,\\
I_2^{disc}(A_2^{max})&:=&\frac{\sqrt{A^{max}_2\Delta'}}{\sqrt{2}(2^{A^{max}_2}-1)}\sum_{1}^{A^{max}_2}\left(\begin{matrix}A^{max}_2 \\ A_2 \end{matrix}\right)\frac{1}{\sqrt{A_2}}\,,
\end{eqnarray}
where we have used \eqref{areamax} and \eqref{binomialesommato}. The sequences $I_1^{disc}(A^{max}_1)$ and $I_2^{disc}(A^{max}_2)$ are plotted in figs. \ref{Fig_I1} and \ref{Fig_I2} together with their continuous approximations \eqref{I1},\eqref{I2}, from which we easily read their maximum values: $max(I^{disc}_1)=0.576=1.414/\sqrt{\Delta'}$ and $max(I^{disc}_2)=2.616=1.065\sqrt{\Delta'}\,.$
 
Thus,
\be
\left|\frac{\partial H}{p_j\partial c_j}\right|\leq\frac{1}{8\pi\gamma^2}\,2\,\left[\max_{A^{max}_1 \geq 1} I_1^{disc}(A_1^{max})\right]^2 \max_{A^{max}_2 \geq 1} I_2^{disc}(A_2^{max})=\frac{2.129}{\sqrt{\Delta'}4\pi \gamma^2}\,.
\ee
From definition \eqref{expansion2} we find the following upper bound for the absolute value of the expansion scalar:
\be
|\theta| \leq 3\,\frac{8\pi\gamma}{2}\, \frac{2.129}{\sqrt{\Delta'}4 \pi\gamma^2}=\frac{6.387}{\sqrt{\Delta'}\gamma}=10.8375\,,\label{bound_teta}
\ee
from \eqref{Bianchishear}, the upper bound for the shear
\be
\sigma^2\leq 3\,\frac{(8\pi\gamma)^2}{3}\,\left(2\,\max_{A^{max}\geq 1}\left|\frac{\partial H}{p_j\partial c_j}\right|\right)^2=\frac{72.522}{\Delta' \gamma^2}=208.7998\,.\label{bound_sig}
\ee

\subsection{$\rho$, $\theta$ and $\sigma^2$ along physical motions}

Here we show the numerical evolutions of the quantities $\rho\,,$ $\theta\,,$ and $\sigma^2$ along the physical motion associated to the KL set of initial conditions discussed before (see tab.\ref{Table_ic}). Their maxima are reported in the captions and they all respect the analytical bounds \eqref{bound_rho},\eqref{bound_teta} and \eqref{bound_sig}. ISO and KUL cases provide similar plots, therefore we do not show them here. In particular, in the QRLG model all the quantities turn out to reach a maximum value that is significantly smaller than the one reached in LQC. Their maximum relative difference between the two models $\Delta \rho:= |\rho_{LQC}-\rho_{QRLG}|/\rho_{LQC},\,$ $\Delta\theta:=|\theta_{LQC}-\theta_{QRLG}|/\theta_{LQC},\,$ $\Delta\sigma^2:=|\sigma^2_{LQC}-\sigma^2_{QRLG}|/\sigma^2_{LQC}$  are 16.76\%, 11.09\% and 28.61\%, respectively.
 
\begin{figure}[H]
	\begin{center}
		\begin{minipage}{0.4\textwidth}
			\centering
			\vspace{0.1cm}
			\includegraphics[width=8.5cm]{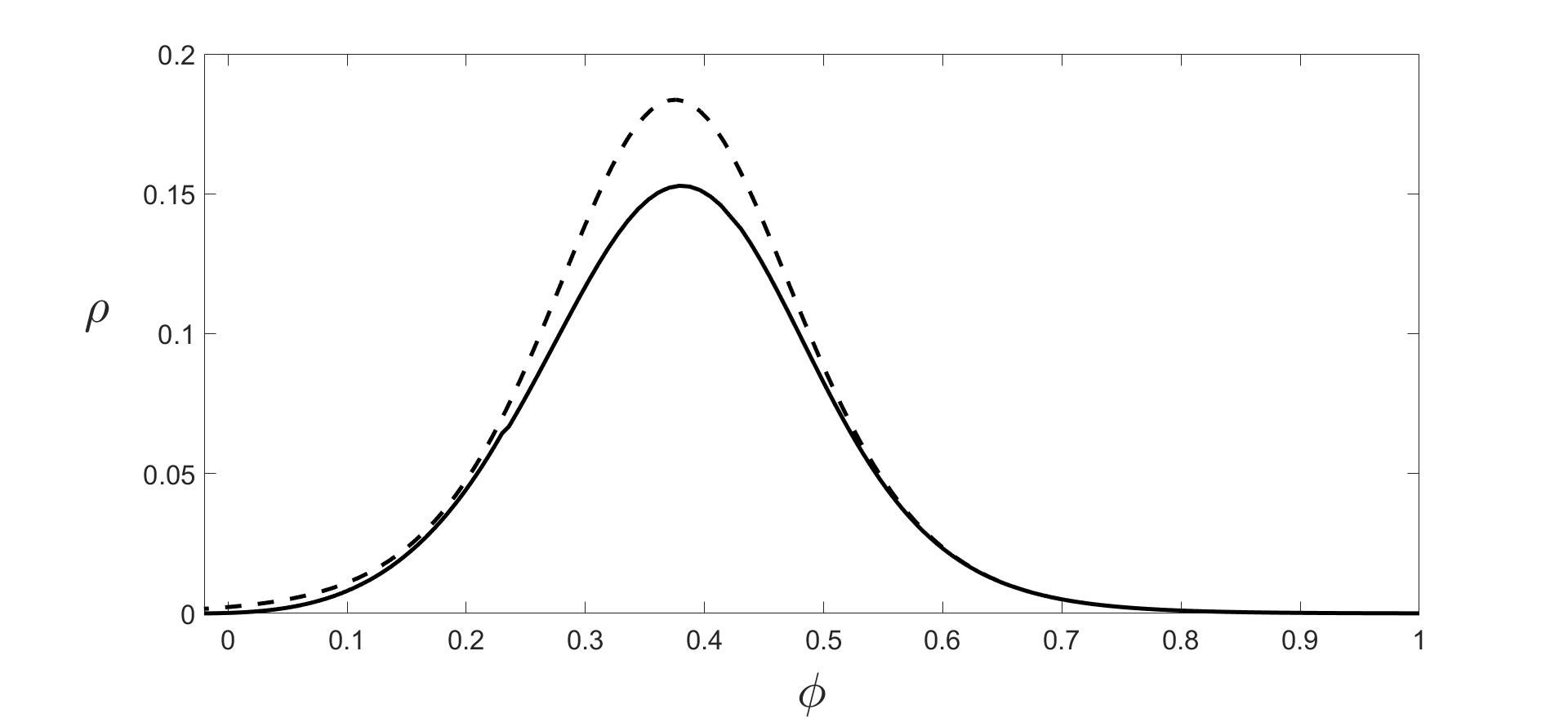}
		\end{minipage}
		\hspace{0.99cm} 
		\begin{minipage}{0.4\textwidth}
			\centering
			\includegraphics[width=8.3cm]{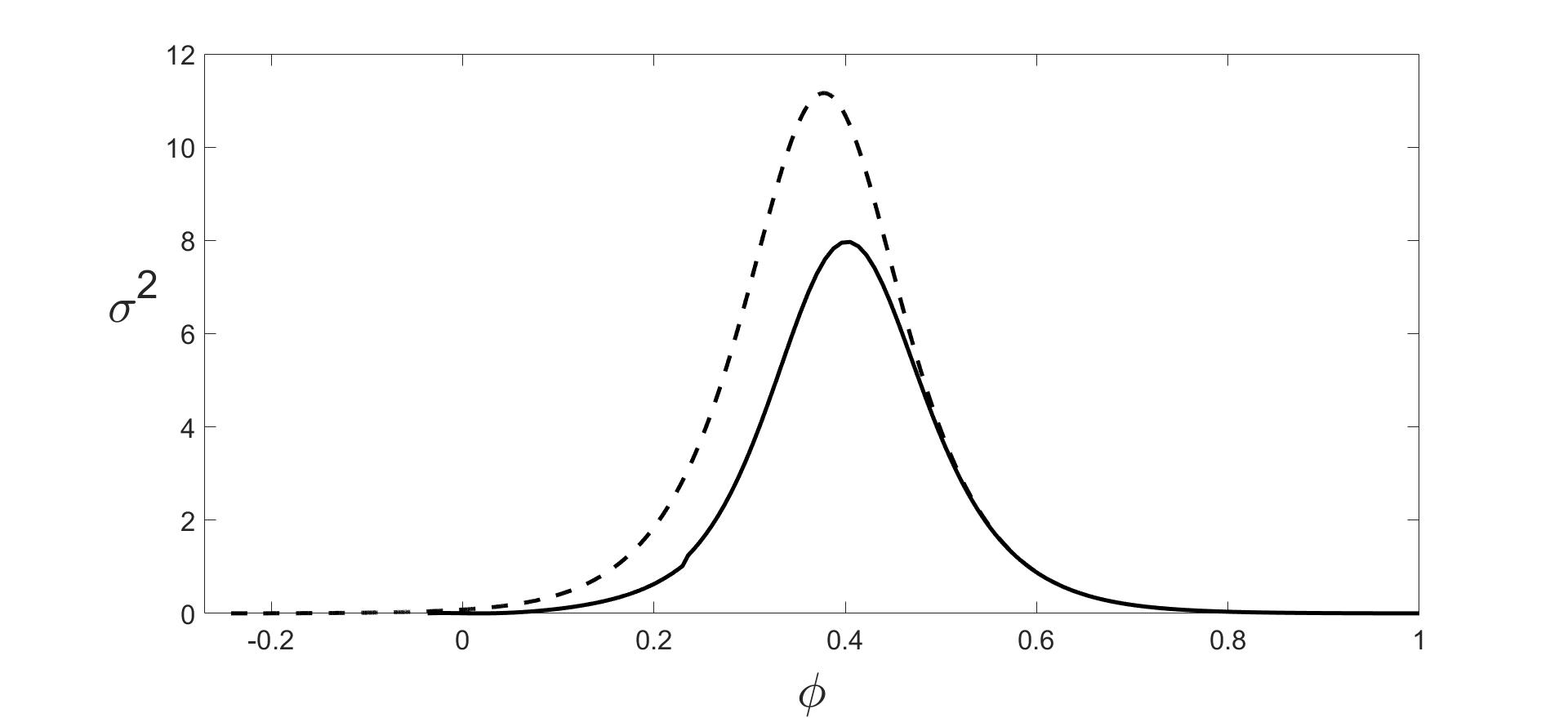}
		\end{minipage}
	\end{center}
	\caption{\small{\emph{Comparison between the QRLG (solid line) and the LQC (dashed line) model for Kasner-like initial conditions. Left panel: evolution of the energy density $\rho$ vs. $\phi$. Right panel: evolution of the shear $\sigma^2$ vs. $\phi$. The maximum and the minimum values are: $\rho^{max}_{QRLG}=0.1529,$ $\rho^{max}_{LQC}=0.1837,$ $\sigma^{2max}_{QRLG}=7.9670,$ and $\sigma^{2max}_{LQC}=11.1605$.}}}\label{fig_rho_sig}
\end{figure}

\begin{figure}[H]
	\begin{center}
		\begin{minipage}{0.4\textwidth}
			\centering
			\vspace{0.1cm}
			\includegraphics[width=8.5cm]{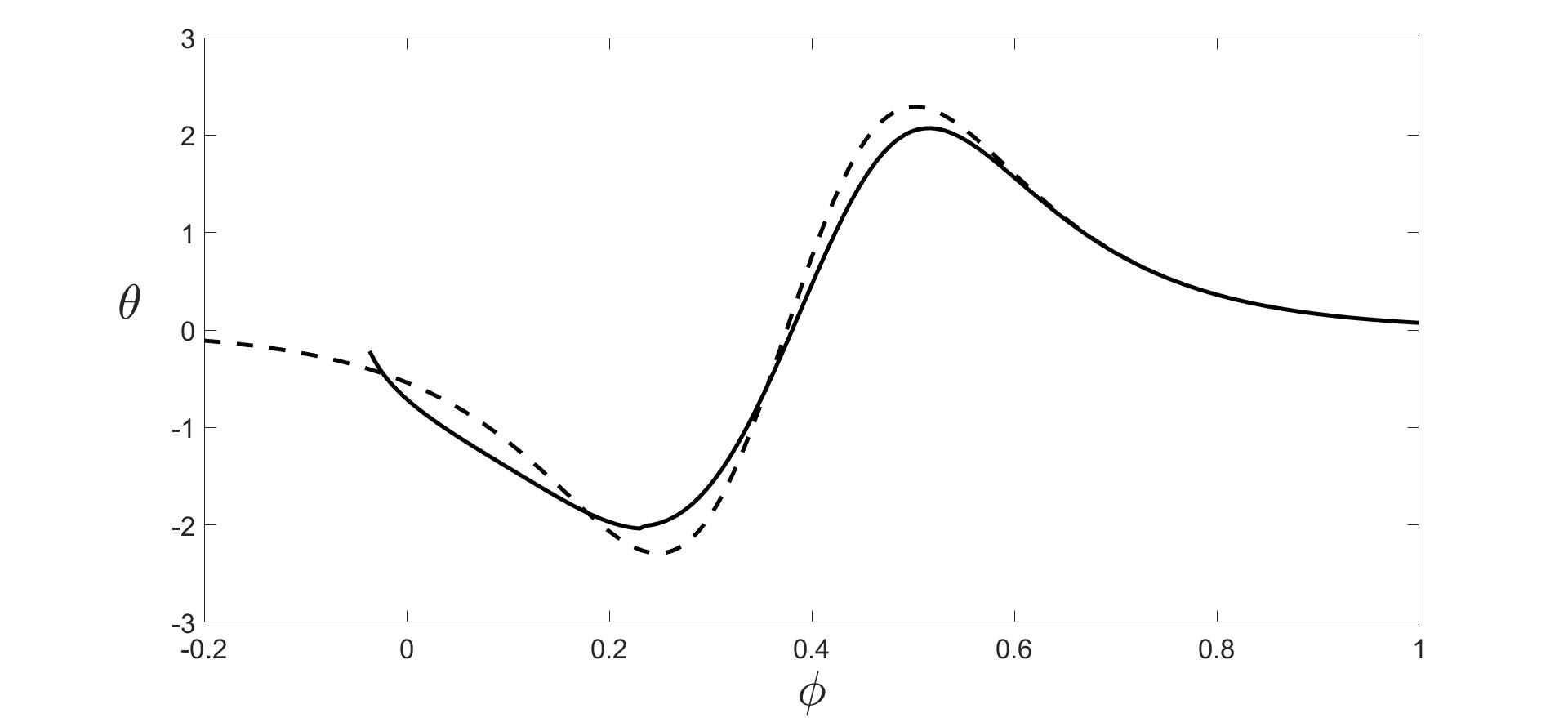}
		\end{minipage}
		\hspace{0.99cm} 
		\begin{minipage}{0.4\textwidth}
			\centering
			\includegraphics[width=8.3cm]{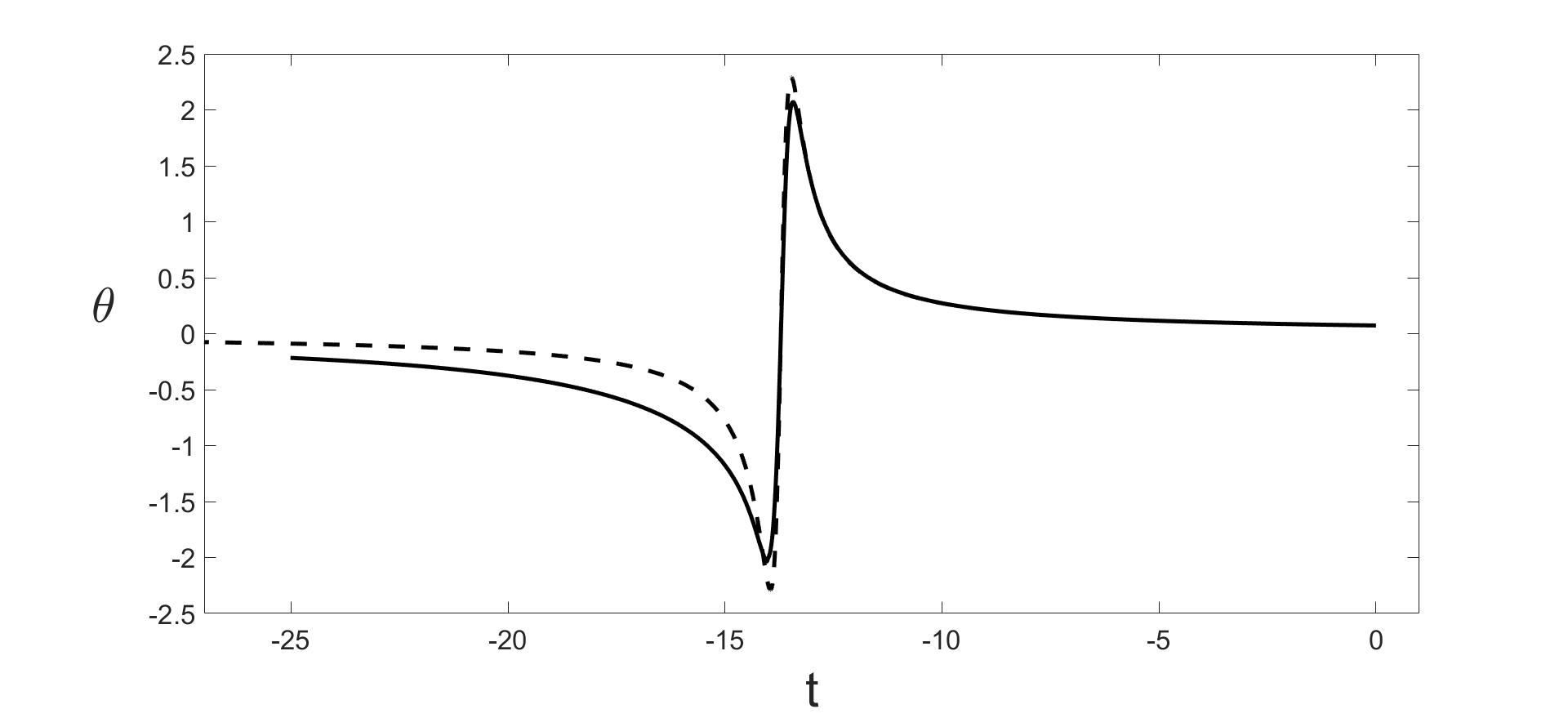}
		\end{minipage}
	\end{center}
	\caption{\small{\emph{Comparison between the QRLG (solid line) and the LQC (dashed line) model for Kasner-like initial conditions. Left panel: evolution of the expansion scalar $\theta$ vs. $\phi$. Right panel: evolution of the expansion scalar $\theta$ vs. $t$. The maximum and the minimum values are: $\theta^{max}_{QRLG}=2.0712,$ $\theta^{max}_{LQC}=2.2909,$ $\theta^{min}_{QRLG}=-2.0368,$ and $\theta^{min}_{LQC}=-2.2909\,.$  }}}\label{fig_teta} 
\end{figure}

Finally, note the evolution of $\theta$ in the relational time $\phi$ (left panel of fig.\ref{fig_teta}) is consistent with the dynamics of $\phi\,,$ plotted in the left panel of fig.\ref{fig_phi_and_mui_cneg}. Indeed, as we approach $\phi=-0.0365\,$,  $d\phi/dt\approx0$ and thus $d\theta/d\phi\equiv (d\theta /dt)\,(d\phi/dt)^{-1}$ speeds up as observed. Moreover, the ``accumulation point" $\phi=-0.0365$ is reached for $t\rightarrow-\infty$ and there is no chance for $\theta$ to grow more than what observed in fig.\ref{fig_teta}.

\section{Conclusions}\label{Sec_conclusions}

 Since a complete theory of quantum gravity is still lacking, insights about the Planck-scale physics relie on several different approaches, such as the one pursued by String theory \cite{Polchinski:1998rq,Polchinski:1998rr}, LQG \cite{Thiemann:2007zz,Rovelli:2004tv}, and Non-commutative geometry \cite{Connes:2000by}, only to cite a few. Even within a given approach, mathematical ambiguities can arise when the machinery of a given theory is applied to a symmetry reduced system to perform computations. This is the case of LQG, where several possible models for the description of our primordial universe come from its covariant \cite{Bianchi:2010zs,Kisielowski:2012yv,Sarno:2018ses}, canonical \cite{Agullo:2016tjh,Dapor:2017rwv,Alesci:2017kzc} and group field theory \cite{Gielen:2016dss} formulation.
 
 In this paper we have studied an anisotropic, homogeneous model for the quantum cosmology of Bianchi I geometry in LQG within the QRLG framework and compared it with effective LQC. The numerical simulations of the evolution of the LQC-Bianchi I and the QRLG-Bianchi I model have all been ran starting from (approximately) common inital conditions that correspond to a classical Bianchi I universe. The LQC evolution is observed to bridge two classical Bianchi I universes before and after the bounces, in agreement with \cite{Chiou:2007mg}, whereas this does not happen for the QRLG-model. Going backwards in the relational time $\phi$, the QRLG evolution starts departing from the LQC one already a bit before the bounces. Those occur only once in each direction, and then an accelerated evolution of each scale factor is observed. The main result is that the QRLG dynamics resolves the classical singularity for \textit{all} kind of initial conditions. In particular, the scale factors turn out to vanish in \textit{all} directions, contrary to what happens in LQC. In the latter case, one of the scale factor turns out to be vanishing in the far past for Kasner-like initial conditions, confirming what already observed in \cite{Chiou:2007mg}. The simulations have been done for three different kinds of classical initial conditions, namely the isotropic, Kasner-like and Kasner-unlike ones. The reliability on the observed singularity resolution in QRLG is strenghtened by the analytical upper bounds we have found for the energy density, expansion scalar and shear.
 
 Another difference between the QRLG-model and the LQC one is that for isotropic initial conditions, the former does not reduce to the QRLG-FLRW model, i.e. to the emergent bouncing universe\cite{Alesci:2016xqa,Alesci:2017kzc}. Instead, a LQC-Bianchi I reduces to a LQC-FLRW. This is clear from the matematical point of view, since a dynamics that starts with isotropic conditions keep evolving isotropically and since the QRLG-Bianchi I Hamiltonian does not reduce to the QRLG-FLRW one in the isotropic limit $p_i\rightarrow p\,,$ $c_i\rightarrow c$. Therefore, the isotropic dynamics of the QRLG-Bianchi I must differ from the ones of the QRLG-FLRW, as observed. Thus, contrary to what happens for LQC, the pre-bounce phenomenological traces of a QRLG primordial universe could be used in principle to understand whether the late (isotropic) universe comes from the isotropic QRLG-Bianchi I model or the QRLG-FLRW one. 
 
 A comparison between the isotropic QRLG-Bianchi I and the recently introduced Dapor-Liegener model \cite{Dapor:2017rwv} comes natural, since both show a departure from standard LQC before the bounce. Backwards in the cosmological time $t$, the latter describes an isotropic universe that starts as a contracting classical FLRW, undergoes a bounce and expands forever according to a non Friedmanian evolution whose limit in the far past is exponential, i.e. driven by a Planckian valued cosmological constant. Thus, both the QRLG-Bianchi I and the Dapor-Liegener model do not exit the quantum regime after the bounce but while the latter here expands exponentially, the former does it only linearly. 

In closing, our study has shown that QRLG offers a viable alternative to the standard picture drawn by LQC, providing a singularity-free model for anisotropic cosmology. Investigations both in the phenomenological and theoretical side remain to be done in the near future, such as computing the power spectrum of tensor and scalar perturbations propagating on the QRLG-Bianchi I effective background, and addressing the quantum dynamics by using a graph-changing Hamiltonian operator. The former will (hopefully) provide crucial observable predictions for testing the effective scenario and the latter deeper insights on QRLG-quantum dynamics beyond its effective scheme.

\section*{Acknowledgements}
This work was supported in part by the NSF grant PHY-1505411, the Eberly research funds of Penn State.

%\newpage
\small\bibliography{BIBLIObianchi}  % Replace xxx by your  usercode (no extension)
\bibliographystyle{unsrt}

\end{document}